\documentclass[aps,prl,twocolumn,floatfix,superscriptaddress,amsmath,amssymb]{revtex4-1}

\usepackage{graphicx}
\usepackage{bm}
\usepackage[pdftex]{color}
\usepackage{xcolor}
\usepackage{hyperref}
\usepackage[mathlines]{lineno}
\usepackage{braket}
\usepackage{bbm}
\usepackage{verbatim}
\usepackage{ulem}
\usepackage{standalone}
\usepackage{comment}
\usepackage[height=8.85in,width=6.45in]{geometry}
\usepackage{slashed}
\usepackage{braket}
\usepackage{tikz}
\usepackage{tikz-cd}
\usepackage{times}
\usepackage{courier}
\usepackage{bm}
\usepackage{xcolor}
\usepackage{mdframed}
\usepackage{datetime}
\usepackage{dsfont}
\usepackage{multirow}
\usepackage{cancel}
\usepackage{caption}
\usepackage{subcaption}
\usepackage{graphicx}
\usepackage[compat=1.1.0]{tikz-feynman}
\usepackage{pifont}
\usepackage{array} 
\usepackage{standalone}

\newcommand{\parR}[1]{\textbf{\textit{#1}}---}

\newcommand{\Z}{\mathbb{Z}}

\renewcommand\[{\begin{equation}}
\renewcommand\]{\end{equation}}

\def\ie{\begin{equation}\begin{aligned}}
\def\fe{\end{aligned}\end{equation}}

\begin{document}

\title{Mixed-State Topological Phase: Quantized Topological Order Parameter and Lieb-Schultz-Mattis Theorem}

\author{Linhao Li}
\affiliation{Department of Physics, the Pennsylvania State University, University Park, Pennsylvania 16802, USA}

\author{Yuan Yao}
\email{smartyao@sjtu.edu.cn}
\thanks{Corresponding author.}
\affiliation{School of Physics and Astronomy, Shanghai Jiao Tong University, Shanghai 200240, China}


\date{\today}

\begin{abstract}
We investigate the extension of pure-state symmetry protected topological phases to mixed-state regime with a strong U(1) and a weak $\mathbb{Z}_2$ symmetries in one-dimensional spin systems by the concept of quantum channels.
We propose a corresponding topological phase order parameter for short-range entangled mixed states by showing that it is quantized and its distinct values can be realized by concrete spin systems with disorders,
sharply signaling phase transitions among them.
We also give a model-independent way to generate two distinct phases by various types of translation and reflection transformations.
These results on the short-range entangled mixed states further enable us to generalize the conventional Lieb-Schultz-Mattis theorem to mixed states,
even without the concept of spectral gaps and lattice Hamiltonians.

\end{abstract}
\maketitle

\parR{Introduction.} 
Distinguishing and characterizing various quantum phases is a central subject in condensed matter and statistical physics.
Symmetry has played an essential role in the phase identifications,
such as the conventional Landau-Ginzburg-Wilson spontaneously symmetry breaking framework ~\cite{Landau:1937aa}.
Significant progress has been made beyond the Landau paradigm,
e.g., the symmetry-protected topological (SPT) phases~\cite{PhysRevB.80.155131,Chen:2010aa,PhysRevB.83.035107,PhysRevB.85.075125,PhysRevB.87.125145,PhysRevB.87.155114} where the ground state can be adiabatically deformed into product states by a finite-depth local unitary circuit (FDLUC) --- so-called short-range entangled (SRE) states \cite{zeng2015quantum},
while such a deformation is forbidden in the presence of certain symmetry. 
Another notable example is the Lieb–Schultz–Mattis (LSM) theorem~\cite{Lieb:1961aa} and its generalizations ~\cite{Affleck:1986aa, OYA1997,Oshikawa:2000aa, Hastings:2004ab,Chen-Gu-Wen_classification2010,Fuji-SymmetryProtection-PRB2016,Watanabe:2015aa,Ogata:2018aa,Ogata:2020aa,Yao:2021aa,Affleck:1986aa,Yao:2019aa,PhysRevB.110.045118,PhysRevB.106.224420,PhysRevLett.133.136705},
which exclude a SPT ground state when the system satisfies nontrivial symmetry data.

The above investigations of quantum phases focus on closed systems and characterize gapped phases by their ground states. 
However,
disorders and decoherence are inevitable in real experiments,
which leaves the systems as mixed states \cite{PhysRevLett.43.1434,PhysRevB.22.1305,PhysRevLett.48.344,PhysRevLett.74.1226,PhysRevLett.103.047201,preskill2018quantum,PhysRevLett.126.130403,PhysRevLett.127.270503,PRXQuantum.3.040313}.
In this context, 
the familiar SRE structure is generalized to density matrices $\rho$.
Correspondingly, the concept of symmetry is also extended;
some symmetry $G$,
called ``weak symmetry'',
may not be respected by each single component of $\rho$,
but it can be averagely preserved by the entire ensemble $g\rho=\rho g,~ \forall g\in G$.
In contrast,
the so-called ``strong symmetry'' $K$ is respected by every component of $\rho$.
Great efforts have been made for the mixed-state topological phases \cite{PhysRevX.14.031044,PhysRevLett.134.070403,yang2025topological,sang2025mixed,PRXQuantum.6.010344,gu2024spontaneous,PRXQuantum.6.010314,PRXQuantum.6.010315,PRXQuantum.6.010313,luo2025topological},
e.g., average SPT (ASPT) phases \cite{Lee_2025,de2022symmetry,zhang2022strange,ma2023average,ma2025topological,PhysRevX.15.021060,xue2024tensor,you2024intrinsic,PRXQuantum.6.020333},
and LSM-type arguments \cite{PhysRevLett.132.070402,PhysRevX.15.011069,PhysRevLett.133.106503,PRXQuantum.6.010347}.
 
Nevertheless,
it is challenging to diagnose and sharply characterize mixed-state SRE (mSRE) phases in the absence of local order parameters.
In analog to pure-state phase classifications,
certain nonlocal string-order parameters may acquire nonzero expectation values in ASPT phases \cite{ma2025topological}.
However, these signals can become arbitrarily small,
preventing a precise sharp distinction among phases. 
Furthermore,
these string-order parameters are strongly model-dependent and there is no universal way to construct a suitable string operator based on $\rho$.
Therefore,
a \textit{model-independent} topological order parameter (i) applicable to topological phases of,
if exists, mixed states and (ii) sharply characterizing distinct phases, e.g., quantized discretely valued, remains an open question.
Finally, 
a rigorous and systematic extension of the LSM theorem,
which forbids mSRE states by certain symmetry constraints,
is still lacking,
partially due to the absence of a commonly accepted definition of “gap’’ for mixed states.

In this letter, we propose the nonlocal order parameter $\text{Tr}(\rho U)$,
where 
$U\equiv\exp\left(\frac{2\pi i}{L}\sum_{j=1}^L jS^z_j\right)$,
to distinguish and characterize mSRE phases in spin-$S$ chains preserving strong $U(1)$ and weak $\Z^x_2$ symmetry.
The so-called twisting operator $U$ was introduced in the proof of LSM theorem for pure states~\cite{Lieb:1961aa}.
Recently,
this order parameter was shown to be sharply quantized in the context of $[U(1)_z\rtimes \Z^x_2]$- SPT classification \cite{PhysRevLett.133.266705}.
We rigorously prove that it can be generalized to mixed-state topological phases,
which is not obvious as it appears since the previous pure-state theorem strongly relies on the lattice Hamiltonian and the gap concept.
Our powerful theorem also enables us to extend LSM theorem to mixed states {with weak (magnetic) translation symmetry or  weak  (magnetic) site-centered reflection symmetry.}

\parR{Preparations and the definitions}We consider a spin-$S$ chain with length $L$ under periodic boundary condition (PBC) and focus on $[U(1)_z \rtimes \mathbb{Z}^x_2]$ spin-rotation symmetry where $U(1)_z$ and $\mathbb{Z}^x_2$ are
generated by $\bm{S}^z_\text{tot}$ and $\pi$-rotation $R^\pi_x$.
\begin{eqnarray}\label{semi_pro}
\bm{S}^z_\text{tot}=\sum^L_{j=1} S^z_n;\,\,\,
R^\pi_x:\,\,R^\pi_x S^z_n\left(R^\pi_x\right)^\dagger=-S^z_n,
\end{eqnarray}
where $R^\pi_x$ can be interpreted as a $\pi$-rotation around $x$-axis.

In the language of Hamiltonians,
it was proven that the unique gapped ground state $|\text{G.S.}\rangle$ of a $[U(1)_z \rtimes \mathbb{Z}^x_2]$-symmetric SPT Hamiltonian satisfies $\langle\text{G.S.}| U|\text{G.S.}\rangle=\pm1+\mathcal{O}(1/L)\rightarrow\pm1$ corresponding to two distinct $[U(1)_z \rtimes \mathbb{Z}^x_2]$-SPT phases \cite{PhysRevLett.133.266705}.

Since there is no Hamiltonian or gap concept for mixed states, we first,
still in the pure-state regime,
extend the above known result so that its real parent Hamiltonian is not referred to,
by the powerful matrix product state (MPS) representations \cite{PhysRevLett.96.220601,perez2006matrix,RevModPhys.93.045003}.

The SRE pure state,
e.g., $|\text{G.S.}\rangle$,
can be represented as an injective MPS with a bond dimension $D$ bounded as $D\leq (2S+1)^{rd}$ when $L$ is sufficiently large~\cite{Fannes:1990ur,Chen:2010aa}.
$r$ is the range of the quantum gates and
$d$ is the circuit depth to deform it to a product state.

{\bf Lemma~1}: For any injective MPS $|\Psi\rangle$
\begin{eqnarray}
\langle s_1,\cdots,s_L|\Psi\rangle=\text{Tr}[A^{[1]}_{s_1}\cdots A^{[L]}_{s_L}],
\end{eqnarray}
which is $G$-symmetric under $\forall g\in G$,
\begin{eqnarray}\label{sym_virtual}
\sum_sU_{s's}(g)A^{[n]}_{s}=\exp[i\alpha_n(g)]V_{n-1}(g)A^{[n]}_{s'}V^\dagger_n(g),
\end{eqnarray}
where $U_{s's}(g)\equiv\langle s'|\hat{U}(g)|s\rangle$ is the unitary physical transformation $\hat{U}(g)$ on the physical degrees of freedom $|s\rangle$,
$\exp(i\alpha_n)$ is the $g$-eigenvalue per site,
and $V_n$ is,
up to a gauge transformation,
the induced effective transformation on the virtual degrees of freedom,
we can construct a $G$-symmetric parent local Hamiltonian $H$ which is gapped with the unique ground state $|\Psi\rangle$,
as a $(k+1)$-neighboring interaction type (where the finite $k$ is the injectivity length \footnote{{Injectivity length $k$ refers to the smallest integer such that the tensors $A^{[n]}_{s_n}\cdots A^{[n+k]}_{s_{n+k}}$ is injective when viewed as a linear transformation $\mathbb{C}^D\rightarrow \mathbb{C}^{d^k}\times\mathbb{C}^D$, where $D$ and $d$ are the dimensions of virtual space and physical space, respectively.}} ): $H=\sum_nh_{n,n+1,\cdots, n+k}$.

\textit{Proof: } We explicitly construct the {Hamiltonian term
$h_{n,n+1,\cdots, n+k}=1-P_{n,n+1,\cdots, n+k}$,
where $P_{n,n+1,\cdots, n+k}$ the orthogonal projection operator of the Hilbert space spanned by states of the form of:
\begin{eqnarray}\label{basis}
\sum_{s_n,\cdots s_{n+k}}\text{Tr}(A^{[n]}_{s_n}\cdots A^{[n+k]}_{s_{n+k}}O)|s_n,\cdots s_{n+k}\rangle,
\end{eqnarray}
where $O$ exhausts all $D\times D$ matrices,
and $h_{n,\cdots,n+k}$ is identity when acted on the Hilbert space of all the other sites (not $n,\cdots,n+k$ sites).}
By the cyclic property of the trace,
the symmetry transformation~(\ref{sym_virtual}) is effectively
\begin{eqnarray}
O\mapsto \exp\left[i\sum_{j=n}^{n+k}\alpha_j(g)\right]V_{n+k}^\dagger (g)O V_{n-1}(g),
\end{eqnarray}
which is bijective on the above Hilbert space spanned by Eq.~(\ref{basis}), thereby leaving $P_{n,n+1,\cdots, n+k}$ and $h_{n,\cdots,n+k}$ $G$-invariant.
It was proved in Ref. \cite{perez2006matrix} that $H$ is indeed gapped with $|\Psi\rangle$ its unique ground state.\hfill$\square$\par

The above explicit construction gives the following Hamiltonian-free Theorem:

{\bf Theorem~2: }For any $[U(1)_z\rtimes\mathbb{Z}^x_2]$-symmetric SRE pure state $|\Psi\rangle$,
we have
\begin{eqnarray}
\langle \Psi| U|\Psi\rangle=\pm1+\mathcal{O}(1/L).
\end{eqnarray}

\parR{Generalization to mixed states}
We are ready to extend the phase characterization from the pure state to the mixed state by density matrices and quantum channels.
We focus on $\rho$ that preserves strong $U(1)_z$ symmetry but only weak $\Z^x_2$ symmetry:
$U(1)_z:\, \exp(i\theta \bm{S}^z_\text{tot})\rho=\exp(i\alpha_\theta)\rho,\,\theta\in[0,2\pi)$ and $\mathbb{Z}_2^x:\,R^{\pi}_x\rho=\rho R^{\pi}_x$,
where $\alpha_\theta=0\text{ or }\pi$ due to $R^\pi_x$ symmetry.
Actually $\alpha_\theta=\alpha_{\theta=0}=0$ by the continuity of the $U(1)_z$ group and then
\begin{eqnarray}\label{u(1)_strong}
\bm{S}^z_\text{tot}\rho=0.
\end{eqnarray}
In this work,
we will use $\Gamma$ to denote the full symmetry group of $\rho$,
which includes both the strong and the weak symmetries.
$U(1)_z$ will be always taken as a strong symmetry,
and $\mathbb{Z}_2^x$ as a weak symmetry.

A natural generalization of FDLUC for the pure states is the finite depth local channel (FDLC) which describe a generic locality-preserving evolution of mixed states.
A general FDLC transformation on any given initial mixed state $\rho_0$ can be constructed in the following steps~[Hastings, 2011]: 
\begin{itemize}
\item (a) Introduce additional qubits as the environment~($E$) on each site $i$ of the system~($S$) to define an enlarged Hilbert space $\mathcal{H}^S_i\otimes \mathcal{H}^E_i$ per site.
The environment is initialized in some product state $|0_E\rangle$. 

\item (b) Apply a FDLUC $D_{S\cup E}$ to the total system $S\cup E$. 

\item (c) Trace out the environment. 
\end{itemize}
Thus a FDLC denoted as $\mathcal{N}$ can be written as
\begin{equation}
\mathcal{N}[\rho_0]=\text{Tr}_E\{D_{S\cup E}\left[\rho_0\otimes|0_E\rangle\langle 0_E|\right]D_{S\cup E}^\dagger\}.
\end{equation}

Moreover, this FDLC is called \textit{locally weak} $\Z^x_2$-symmetric~\cite{Chen:2010aa,PRXQuantum.6.010347},
if (i) $|0_E\rangle$ is a $\Z^x_2$-symmetric product state, $R_E|0\rangle_E\propto|0\rangle_E$,
where $R_E$ is the $\Z^x_2$ generator on the environment,
and
(ii) the layer decomposition $D_{S\cup E}=\prod_{j=1}^{n_\text{layer}}D^{(j)}_{S\cup E}$ satisfies $[D^{(j)}_{S\cup E},R^{\pi}_x\otimes R_E]=0$ for each layer $j$.
Similarly,
e.g.,
for a \textit{locally strong} U(1)$_z$-symmetric FDLC,
$R_E$ in (i) is replaced by the environment U(1)$_z$ operator $U^{\theta}_E$ with $\theta\in[0,2\pi)$, and
$R^\pi_x\otimes R_E$ in (ii) by $\exp(i\theta\bm{S}^z_\text{tot})\otimes 1_E$ with $1_E$ the environment identity operator.
FDLC is also symmetry preserving;
$\mathcal{N}[\rho_0]$ is $\Gamma$-symmetric if $\rho_0$ is $\Gamma$-symmetric and $\mathcal{N}$ is locally $\Gamma$-symmetric.

As FDLC generalizes the concept of FDLUC,
we have the following extension of SRE for mixed states:

{\bf Definition: }$\rho$ is a {\bf $\Gamma'$-mSRE state}, if $\rho=\mathcal{N}[\otimes_j\rho_j]$,
where $\mathcal{N}$ is a locally $\Gamma'$-symmetric FDLC and $\otimes_j\rho_j$ is a product of $\Gamma'$-symmetric mixed states $\rho_j$ each of which can include several lattice sites around $j$ coordinates forming a $j$-``cell'' and the cells in the product $\otimes_j\rho_j$ have no overlap with each other.

Generally,
the symmetry $\Gamma$ of a $\Gamma'$-mSRE $\rho$ can be strictly larger than $\Gamma'$;
$\Gamma'$ reflects how the $\rho$ is prepared while $\Gamma$ is the symmetry of the produced state $\rho$.
Now,
we are ready to present our main result:

{\bf Theorem~3: }
If $\rho$ respects strong $U(1)_z$-symmetry and weak $\mathbb{Z}_2^x$-symmetry,
then
\begin{equation}
\text{Tr}(\rho U)=\pm 1+\mathcal{O}(1/L),
\end{equation}
as long as $\rho$ is (weak $\mathbb{Z}^x_2$)-mSRE state.

{\bf Remark:} $\rho$ needs to be a (weak $\mathbb{Z}_2$)-mSRE state rather than a general mSRE state,
in contrast to its pure-state version as {\bf Theorem~4}.
The extension to mixed states requires this nontrivial enriched symmetry structure.

\textit{Proof: }
Following Eqs.~(27-29) of Ref.\cite{PRXQuantum.6.010347} which proves that
any (weak $\mathbb{Z}_2^x$)-symmetric product mixed state can be purified by introduction of ancilla (A):
\begin{equation}\label{puri_1}
\otimes_j\rho_j=\text{Tr}_A(|\psi_{S\cup A}\rangle\langle\psi_{S\cup A}|)
\end{equation}
where an ancilla local Hilbert space $\mathcal{H}^j_A$ is a copy of the original spin Hilbert space $\mathcal{H}^j_S$ at each cell $j$.
The state $|\psi_{S\cup A}\rangle$ is a product state satisfying
\begin{equation}
[R^{\pi}_x \otimes (R^\pi_x)^*]|\psi_{S\cup A}(L)\rangle = |\psi_{S\cup A}(L)\rangle,
\end{equation}
where the complex conjugate $(R^\pi_x)^*$ is acting on the ancilla.
Recalling $\rho=\mathcal{N}[\otimes\rho_j]=\text{Tr}_E\{D_{S\cup E}\left[(\otimes_j\rho_j)\otimes|0_E\rangle\langle 0_E|\right]D_{S\cup E}^\dagger\}$ and Eq.~(\ref{puri_1}),
we obtain the following purification
\begin{eqnarray}\label{puri_2}
&&\rho=\text{Tr}_{E,A}(|\Phi\rangle\langle\Phi|),\\
&&|\Phi\rangle\equiv(D_{S\cup E}\otimes \mathbb{I}_A)|\psi_{S\cup A}\rangle\otimes |0\rangle_E,\label{SRE_Phi}
\end{eqnarray}
where $\mathbb{I}_A$ is the identity on the ancilla.
Obviously,
the pure state 
$|\Phi\rangle$ defined in the total Hilbert space $\mathcal{H}^S \otimes \mathcal{H}^A \otimes \mathcal{H}^E$ is invariant under the $\mathbb{Z}^x_2$ symmetry with representation:
\begin{equation}
R^x_{\pi}\otimes (R^\pi_x)^*\otimes R_E|\Phi\rangle=|\Phi\rangle.
\end{equation}
It also satisfies the following extended $U(1)_z$ symmetry ~\footnote{This can be easily verified by performing the Schmidt decomposition of $|\Phi\rangle$ under the bipartition $S\cup AE$}:
\begin{equation}
\exp(i\theta S^z)\otimes \mathbb{I}_{E\cup A}|\Phi\rangle=|\Phi\rangle,
\end{equation}
where the $U(1)_z$ is \textit{defined} to be trivial on the artificial ancilla and environment,
which is consistent with $R^\pi_x S^z_n\left(R^\pi_x\right)^\dagger=-S^z_n$, where $S^z_{E,n}=S^z_{A,n}\equiv0$.
By Eq.~(\ref{SRE_Phi}),
$\Phi$ is an SRE pure state,
and {\bf Theorem~2} gives
\begin{eqnarray}
\langle\Phi|U_{S\cup A\cup E}|\Phi\rangle=\pm1+\mathcal{O}(1/L),
\end{eqnarray}
where
\begin{eqnarray}\label{twist_tot}
U_{S\cup A\cup E}&\equiv&\exp\left[\frac{2\pi i}{L}\sum_{n=1}^Ln(S^z_n+S^z_{A,n}+S^z_{E,n})\right]\nonumber\\
&=&U\otimes\mathbb{I}_{E\cup A}.
\end{eqnarray}

Due to Eqs.~(\ref{puri_2},\ref{twist_tot}),
we obtain
\begin{equation}
\begin{split}
\text{Tr}(\rho U)&=\text{Tr}_{S\cup A\cup E}\left(|\Phi\rangle\langle\Phi|U_{S\cup A\cup E}]\right)\\&=\langle\Phi|U_{S\cup A\cup E}|\Phi\rangle\\&=\pm 1+\mathcal{O}(1/L),
\end{split}
\end{equation}
which completes the proof of {\bf Theorem~3}.\hfill{$\square$}

\parR{Mixed-state topological phase and topological order parameter}
The classification of topological phases can be extended to the mixed-state regime as follows.

{\bf Definition:} (Mixed-state topological phase) Two density $\Gamma$-symmetric matrices $\rho_1$, $\rho_2$ are said to belong to the same $\Gamma$-symmetric phase,
called $\Gamma$-symmetric phase,
if they are two-way connected by a pair of locally $\Gamma$-symmetric FDLCs ~\cite{PhysRevX.14.031044}

The discretely valued $\langle U\rangle=\text{Tr}(\rho U)$ as $L\rightarrow\infty$ implies that it can be used to distinguish at least two distinct mixed-state topological phases.
It can be seen by the following continuity consideration.
Let us assume there is a FDLC connecting $\rho_\text{in}$ and $\rho_\text{out}$ satisfying the condition of {\bf Theorem~3}.
In the aforementioned general construction step~(b) of FDLC,
the FDLUC therein can be simulated with a continuous unitary evolution by a local Hamiltonian for a finite time $t\in[0,1]$~\cite{Chen:2010aa}.
Therefore,
if this unitary evolution is from $t=0$ only until $t=s$ followed by the step~(c) of tracing the environment out,
we have a continuous $s$-family~\footnote{It is generally not $s$-differentiable for multi-layer cases.} of FDLCs and thereby a $s$-parameterized density matrices $\rho_s$ with $\rho_{s=0}=\rho_\text{in}$ and $\rho_{s=1}=\mathcal{N}[\rho_\text{in}]=\rho_\text{out}$.
The simulating unitary evolution until any $s$ also preserves $[U(1)_z\rtimes\mathbb{Z}_2^x]$ symmetry since the simulating local Hamiltonian,
e.g., $i\log(D^{(j)}_{S\cup E})$ for each layer $j$,
preserves $[U(1)_z\rtimes\mathbb{Z}_2^x]$.
Thus,
$\rho_s$ satisfies the condition of {\bf Theorem~3} for any $s\in[0,1]$, which enables us to calculate and conclude that
\begin{eqnarray}
\mathcal{I}_s\equiv\lim_{L\rightarrow\infty}\text{Tr}(\rho_sU)\in\{\pm1\},
\end{eqnarray}
which must be a constant function about $s$ since it is discretely $(\pm1)$-valued and $s$-continuous.
Particularly,
$\mathcal{I}_{s=1}=\mathcal{I}_0$,
which shows that,
conversely,
different values of $\mathcal{I}\equiv\lim_{L\rightarrow\infty}\langle U\rangle$,
as a quantized topological order parameter,
sharply signal distinct mixed-state matter of phases \footnote{We note that a refined definition of mixed-state phase equivalence—based on one-way connectivity via locally reversible channel circuits—was recently introduced in Ref. \cite{sang2025mixed}, motivated by the existence of a counterexample in two-dimensional classical statistical mechanics under the original two-way connectivity definition. It has been shown in \cite{PhysRevLett.134.070403} that two states being equivalent under this refined definition implies their equivalence under two-way connectivity. Thus two states with distinct values of $\mathcal{I}$ must also belong to distinct mixed-state matter of phases even under this refined definition. 
}.

\begin{figure}[h]
\centering
{\includegraphics[width=0.5\textwidth]{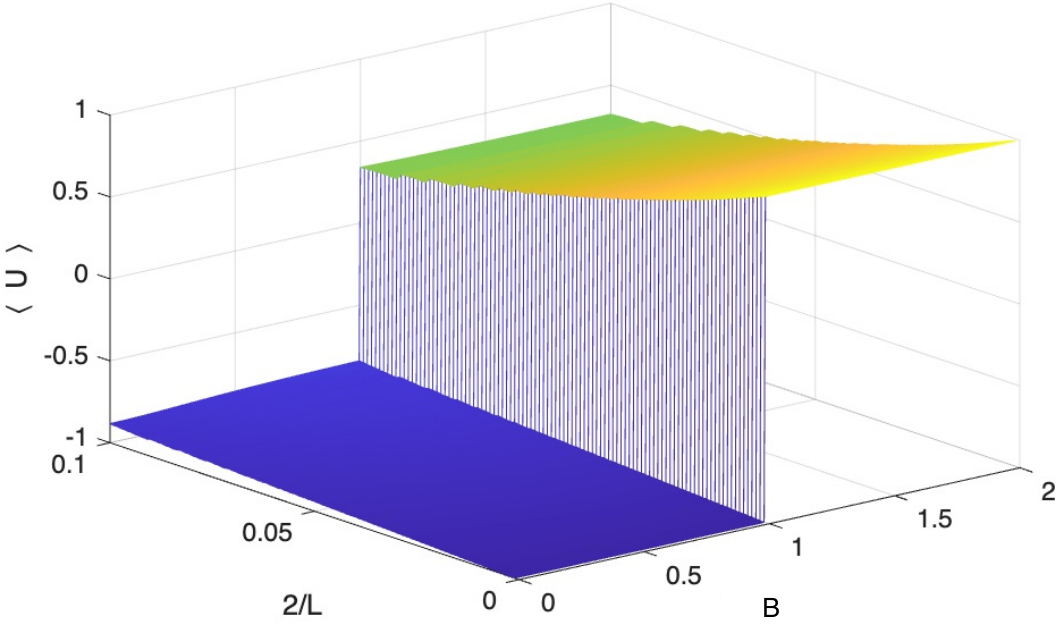}}
\caption{Numerical calculation of $\langle U\rangle$ with varying $L$ and $B$ identifies two topological phases: (i) $\mathcal{I}=-1$ when $0\leq B<1$, and (ii) $\mathcal{I}=+1$ when $B>1$.}
\label{fig:phase_transition}
\end{figure}

Let us confirm our rigorous statement by illustrating examples.
Both $\mathcal{I}=\pm1$ can be indeed realized on lattices by the following exactly solvable model.
The clean Hamiltonian is a spin-1/2 chain $H_0=-\sum_{n=1}^N\vec{S}_{2n}\cdot\vec{S}_{2n+1}$,
and the disorder is
\begin{eqnarray}
\Delta H(B)=\sum_{n=1}^N2h_n(-S^z_{2n+1}+S^z_{2n+2}),
\end{eqnarray}
where $h_n$ are independent random variable $(B\geq0)$:
\begin{eqnarray}
P(h_n=B)=50\%,\,\,P(h_n=-B)=50\%.
\end{eqnarray}
By a transferred matrix technique~\cite{supplemental},
we find two phases separated by $B_c=1$: [up to $\mathcal{O}(1/L)$]
\begin{eqnarray}\label{ana_disorder}
\langle U\rangle=\left\{\begin{array}{cc}-1+\frac{\pi^2}{8L}(2+\Delta_B^2),&0\leq B<B_c;\\
1-\frac{\pi^2}{8L}(\Delta_B^2+8\Delta_B+9),&B>B_c,
\end{array}\right.
\end{eqnarray}
where $\Delta_B\equiv\frac{4B}{1+\sqrt{1+4B^2}+\frac{4B^2}{1+\sqrt{1+4B^2}}}$.
Two states at $B=0$ and $B\rightarrow\infty$ are both (weak $\mathbb{Z}_2^x$)-mSRE~\cite{supplemental},
and we find two mixed-state realizations predicted by {\bf Theorem~3}.
Although Eq.~(\ref{ana_disorder}) is analytically true,
a direct numerical simulation in FIG.~\ref{fig:phase_transition} illustrates the sharp topological phase transition at $B=B_c$.

Actually,
there is a model-free understanding of existence of two phases by various translation symmetries as shown below.

\parR{Lieb-Schultz-Mattis (LSM) theorem for the mixed state} We observe that both two distinct mixed-state phases above fail to preserve weak translation symmetry.
Actually,
the translation symmetry can be shown to ``connect'' distinct phases:

{\bf Theorem~4: }If a half-integral spin chain is in a (weak $\mathbb{Z}_2$)-mSRE state $\rho$ respects strong $U(1)_z$-symmetry and weak $\mathbb{Z}_2^x$-symmetry, 
then
\begin{equation}
\begin{split}
\mathcal{I}[\rho]&=-\mathcal{I}[T\rho T^{-1}]=-\mathcal{I}[\tilde{T}\rho \tilde{T}^{-1}]\\&=-\mathcal{I}[R\rho R^{-1}]=-\mathcal{I}[\tilde{R}\rho \tilde{R}^{-1}].
\end{split}
\end{equation}
Here $T$ is the lattice translation symmetry 
and $R$ is the site-centered reflection symmetry. $\tilde{T}$ is a general magnetic translation and $\tilde{R}$ is a general magnetic site-centered reflection :
\begin{equation}
\Tilde{T}=T \circ Q \circ \Theta,\quad
    \Tilde{R}=R \circ Q \circ \Theta,
\end{equation}
where $\Theta$ is the conventional time-reversal and $Q$ is any unitary onsite operator that commutes with $S^j_z$. 

\textit{Proof:} 
After applying the translation operation to $\rho$, we obtain
\begin{equation}\label{eq:translation}
\begin{split}
    \text{Tr}(T\rho T^{-1} U)&=\text{Tr}\left(\rho U \exp(2\pi i S^z_1)\exp(-\frac{2\pi i}{L} \bm{S}^z_\text{tot})\right)\\&=(-1)^{2S}\text{Tr}(\rho U)=-\text{Tr}(\rho U)
    \end{split}
\end{equation}
where we have used Eq.~(\ref{u(1)_strong}) that $\bm{S}^z_\text{tot}\rho=0$ due to the weak $\mathbb{Z}_2^x$ symmetry. 

Next, without loss of generality, we consider a reflection $R$ centered at site $L/2$ \footnote{The length $L$ must be even due to $\bm{S}^z_\text{tot}\rho=0$.}. This yields
\begin{equation}
\begin{split}
    \text{Tr}(R\rho R^{-1} U)&=\text{Tr}\left(\rho U^{\dagger} \exp(2\pi i S^z_L)\right)\\&=(-1)^{2S}\text{Tr}(\rho U^{\dagger})\\&=-\text{Tr}(\rho U),
    \end{split}
\end{equation}
where we have used that $\rho$ respects weak $\mathbb{Z}_2^x$ symmetry to deduce $\text{Tr}(\rho U^\dagger)=\text{Tr}(\rho U)$.

The translation transformation $T$ and reflection $R$ above can be replaced by $\tilde{T}$ and 
 $\tilde{R}$ respectively with no substantial change to the proof. \hfill{$\square$}

{\bf Theorem~4} directly gives a realization of two phases once we are given one state $\rho$ in whichever phase. 
It also naturally implies the following mixed-state version of the Lieb–Schultz–Mattis (LSM) theorem:

{\bf Corollary~5: }\textit{(Mixed-state LSM theorem)} 
A mixed state $\rho$ of a spin chain respects strong $U(1)_z$-symmetry,
weak $\mathbb{Z}_2^x$-symmetry and {weak (magnetic) translation symmetry or  weak  (magnetic) site-centered reflection symmetry.} 
Then
it cannot be a weak $\mathbb{Z}_2^x$-mSRE state, if the spin chain has half-integral spin per unit cell. \hfill{$\square$}

{\bf Corollary~5} is thus an extension of the LSM theorem not only to mixed states but also to translation symmetry beyond the unitary case. The original LSM proof relies on the unitarity of translation to define a well-defined lattice momentum, which is unavailable in the antiunitary case. Furthermore, compared with Ref. \cite{PRXQuantum.6.010347}, which also discusses the LSM theorem for mixed states, our result does not require the state to be a (weak $T$)-mSRE state and includes the case of magnetic translations {and (magnetic) reflections},
so it is strictly stronger than the result in Ref. \cite{PRXQuantum.6.010347}.



\parR{Illustrating examples of the mixed-state LSM theorem} Let us explicitly see a mixed state model as an application of the LSM theorem.
We consider the 1d chiral scalar triple-product spin model:
\begin{equation}
\mathcal{H} _\text{ch} = \sum_{r}(-1)^i J\vec{S}_i\cdot(\vec{S}_{i+1}\times\vec{S}_{i+2}).
\label{eq:CTP}
\end{equation}
with ground state $|\text{G.S.}\rangle$ and the channel $\forall p\in[0,1]$
\begin{eqnarray}
    \mathcal{N}=\otimes_i \mathcal{N}^z_i, \quad \mathcal{N}^z_i[\rho]=(1-p)\rho+4 p S^z_i\rho S^z_i
\end{eqnarray}
The Hamiltonian and the channel preserve the strong-$U(1)_z$ and weak-$\Z^x_2$ symmetry since $S^z_i$ in the channel commutes with $U(1)_z$ symmetry and anticommutes with $R^{x}_{\pi}$. 
Moreover, this model also preserves a magnetic translation symmetry:
\begin{eqnarray}
(T\circ\Theta)\vec{S}_r=-\vec{S}_{r+1}(T\circ\Theta).
\end{eqnarray}
Therefore,
{\bf Corollary~5} implies that strong $U(1)_z$- and weak $\mathbb{Z}_2$-symmetric $\mathcal{N}[|\text{G.S.}\rangle\langle\text{G.S.}|]$ cannot be a weak $\mathbb{Z}_2$-mSRE state.
Actually, 
there is a model-dependent proof~\cite{supplemental}, where we show the spin–spin correlation function decays as a power law.

\parR{Conclusions and discussions}
We propose a quantized topological phase order parameter to distinguish different mSRE states and such an order parameter is model-independent and can sharply detect various topological phase transitions,
for which we present an exactly solvable model.
Our results enable us to extend the LSM theorem to mixed states without the concept of spectral gaps.
The generalization to SU(N) spin systems is also straightforward and natural \cite{PhysRevLett.133.266705}.

As another application of {\bf Corollary~5} for mSRE state, we can conclude that
if an mSRE state $\rho$ respects strong $U(1)_z$-symmetry,
weak $\mathbb{Z}_2^x$-symmetry  and {weak (magnetic) translation symmetry or  weak  (magnetic) site-centered reflection symmetry.} 
then $\rho$ cannot be a (weak $\mathbb{Z}^x_2$)-mSRE.
Thus,
it must be in a nontrivial weak-$\mathbb{Z}^x_2$ symmetric topological phase if the spin chain has half-integral spin per unit cell,
since
there is no two-way locally (weak $\mathbb{Z}^x_2$)-symmetric FDLCs connecting $\rho$ and a $\mathbb{Z}_2^x$-symmetric product mixed state.
A lattice construction of such a nontrivial phase along this direction is left for future work.

\parR{Acknowledgments.}
The authors thank Akira Furusaki for useful discussions.
The work of Y. Y. was supported by the National Key Research and Development Program of China (Grant No. 2024YFA1408303), the National Natural Science Foundation of China (Grants No. 12474157 and No. 12447103), the sponsorship from Yangyang Development Fund, and Xiaomi Young Scholars Program.


\begin{thebibliography}{67}%
\makeatletter
\providecommand \@ifxundefined [1]{%
 \@ifx{#1\undefined}
}%
\providecommand \@ifnum [1]{%
 \ifnum #1\expandafter \@firstoftwo
 \else \expandafter \@secondoftwo
 \fi
}%
\providecommand \@ifx [1]{%
 \ifx #1\expandafter \@firstoftwo
 \else \expandafter \@secondoftwo
 \fi
}%
\providecommand \natexlab [1]{#1}%
\providecommand \enquote  [1]{``#1''}%
\providecommand \bibnamefont  [1]{#1}%
\providecommand \bibfnamefont [1]{#1}%
\providecommand \citenamefont [1]{#1}%
\providecommand \href@noop [0]{\@secondoftwo}%
\providecommand \href [0]{\begingroup \@sanitize@url \@href}%
\providecommand \@href[1]{\@@startlink{#1}\@@href}%
\providecommand \@@href[1]{\endgroup#1\@@endlink}%
\providecommand \@sanitize@url [0]{\catcode `\\12\catcode `\$12\catcode
  `\&12\catcode `\#12\catcode `\^12\catcode `\_12\catcode `\%12\relax}%
\providecommand \@@startlink[1]{}%
\providecommand \@@endlink[0]{}%
\providecommand \url  [0]{\begingroup\@sanitize@url \@url }%
\providecommand \@url [1]{\endgroup\@href {#1}{\urlprefix }}%
\providecommand \urlprefix  [0]{URL }%
\providecommand \Eprint [0]{\href }%
\providecommand \doibase [0]{http://dx.doi.org/}%
\providecommand \selectlanguage [0]{\@gobble}%
\providecommand \bibinfo  [0]{\@secondoftwo}%
\providecommand \bibfield  [0]{\@secondoftwo}%
\providecommand \translation [1]{[#1]}%
\providecommand \BibitemOpen [0]{}%
\providecommand \bibitemStop [0]{}%
\providecommand \bibitemNoStop [0]{.\EOS\space}%
\providecommand \EOS [0]{\spacefactor3000\relax}%
\providecommand \BibitemShut  [1]{\csname bibitem#1\endcsname}%
\let\auto@bib@innerbib\@empty
\bibitem [{\citenamefont {Landau}(1937)}]{Landau:1937aa}%
  \BibitemOpen
  \bibfield  {author} {\bibinfo {author} {\bibfnamefont {L.~D.}\ \bibnamefont
  {Landau}},\ }\href {\doibase 10.1016/B978-0-08-010586-4.50034-1} {\bibfield
  {journal} {\bibinfo  {journal} {Zh. Eksp. Teor. Fiz.}\ }\textbf {\bibinfo
  {volume} {7}},\ \bibinfo {pages} {19} (\bibinfo {year} {1937})}\BibitemShut
  {NoStop}%
\bibitem [{\citenamefont {Gu}\ and\ \citenamefont
  {Wen}(2009)}]{PhysRevB.80.155131}%
  \BibitemOpen
  \bibfield  {author} {\bibinfo {author} {\bibfnamefont {Z.-C.}\ \bibnamefont
  {Gu}}\ and\ \bibinfo {author} {\bibfnamefont {X.-G.}\ \bibnamefont {Wen}},\
  }\href {\doibase 10.1103/PhysRevB.80.155131} {\bibfield  {journal} {\bibinfo
  {journal} {Phys. Rev. B}\ }\textbf {\bibinfo {volume} {80}},\ \bibinfo
  {pages} {155131} (\bibinfo {year} {2009})}\BibitemShut {NoStop}%
\bibitem [{\citenamefont {Chen}\ \emph {et~al.}(2010)\citenamefont {Chen},
  \citenamefont {Gu},\ and\ \citenamefont {Wen}}]{Chen:2010aa}%
  \BibitemOpen
  \bibfield  {author} {\bibinfo {author} {\bibfnamefont {X.}~\bibnamefont
  {Chen}}, \bibinfo {author} {\bibfnamefont {Z.-C.}\ \bibnamefont {Gu}}, \ and\
  \bibinfo {author} {\bibfnamefont {X.-G.}\ \bibnamefont {Wen}},\ }\href
  {\doibase 10.1103/PhysRevB.82.155138} {\bibfield  {journal} {\bibinfo
  {journal} {Phys. Rev. B}\ }\textbf {\bibinfo {volume} {82}},\ \bibinfo
  {pages} {155138} (\bibinfo {year} {2010})}\BibitemShut {NoStop}%
\bibitem [{\citenamefont {Chen}\ \emph
  {et~al.}(2011{\natexlab{a}})\citenamefont {Chen}, \citenamefont {Gu},\ and\
  \citenamefont {Wen}}]{PhysRevB.83.035107}%
  \BibitemOpen
  \bibfield  {author} {\bibinfo {author} {\bibfnamefont {X.}~\bibnamefont
  {Chen}}, \bibinfo {author} {\bibfnamefont {Z.-C.}\ \bibnamefont {Gu}}, \ and\
  \bibinfo {author} {\bibfnamefont {X.-G.}\ \bibnamefont {Wen}},\ }\href
  {\doibase 10.1103/PhysRevB.83.035107} {\bibfield  {journal} {\bibinfo
  {journal} {Phys. Rev. B}\ }\textbf {\bibinfo {volume} {83}},\ \bibinfo
  {pages} {035107} (\bibinfo {year} {2011}{\natexlab{a}})}\BibitemShut
  {NoStop}%
\bibitem [{\citenamefont {Pollmann}\ \emph {et~al.}(2012)\citenamefont
  {Pollmann}, \citenamefont {Berg}, \citenamefont {Turner},\ and\ \citenamefont
  {Oshikawa}}]{PhysRevB.85.075125}%
  \BibitemOpen
  \bibfield  {author} {\bibinfo {author} {\bibfnamefont {F.}~\bibnamefont
  {Pollmann}}, \bibinfo {author} {\bibfnamefont {E.}~\bibnamefont {Berg}},
  \bibinfo {author} {\bibfnamefont {A.~M.}\ \bibnamefont {Turner}}, \ and\
  \bibinfo {author} {\bibfnamefont {M.}~\bibnamefont {Oshikawa}},\ }\href
  {\doibase 10.1103/PhysRevB.85.075125} {\bibfield  {journal} {\bibinfo
  {journal} {Phys. Rev. B}\ }\textbf {\bibinfo {volume} {85}},\ \bibinfo
  {pages} {075125} (\bibinfo {year} {2012})}\BibitemShut {NoStop}%
\bibitem [{\citenamefont {Duivenvoorden}\ and\ \citenamefont
  {Quella}(2013)}]{PhysRevB.87.125145}%
  \BibitemOpen
  \bibfield  {author} {\bibinfo {author} {\bibfnamefont {K.}~\bibnamefont
  {Duivenvoorden}}\ and\ \bibinfo {author} {\bibfnamefont {T.}~\bibnamefont
  {Quella}},\ }\href {\doibase 10.1103/PhysRevB.87.125145} {\bibfield
  {journal} {\bibinfo  {journal} {Phys. Rev. B}\ }\textbf {\bibinfo {volume}
  {87}},\ \bibinfo {pages} {125145} (\bibinfo {year} {2013})}\BibitemShut
  {NoStop}%
\bibitem [{\citenamefont {Chen}\ \emph {et~al.}(2013)\citenamefont {Chen},
  \citenamefont {Gu}, \citenamefont {Liu},\ and\ \citenamefont
  {Wen}}]{PhysRevB.87.155114}%
  \BibitemOpen
  \bibfield  {author} {\bibinfo {author} {\bibfnamefont {X.}~\bibnamefont
  {Chen}}, \bibinfo {author} {\bibfnamefont {Z.-C.}\ \bibnamefont {Gu}},
  \bibinfo {author} {\bibfnamefont {Z.-X.}\ \bibnamefont {Liu}}, \ and\
  \bibinfo {author} {\bibfnamefont {X.-G.}\ \bibnamefont {Wen}},\ }\href
  {\doibase 10.1103/PhysRevB.87.155114} {\bibfield  {journal} {\bibinfo
  {journal} {Phys. Rev. B}\ }\textbf {\bibinfo {volume} {87}},\ \bibinfo
  {pages} {155114} (\bibinfo {year} {2013})}\BibitemShut {NoStop}%
\bibitem [{\citenamefont {Zeng}\ \emph {et~al.}(2015)\citenamefont {Zeng},
  \citenamefont {Chen}, \citenamefont {Zhou},\ and\ \citenamefont
  {Wen}}]{zeng2015quantum}%
  \BibitemOpen
  \bibfield  {author} {\bibinfo {author} {\bibfnamefont {B.}~\bibnamefont
  {Zeng}}, \bibinfo {author} {\bibfnamefont {X.}~\bibnamefont {Chen}}, \bibinfo
  {author} {\bibfnamefont {D.-L.}\ \bibnamefont {Zhou}}, \ and\ \bibinfo
  {author} {\bibfnamefont {X.-G.}\ \bibnamefont {Wen}},\ }\href
  {https://arxiv.org/abs/1508.02595} {\bibfield  {journal} {\bibinfo  {journal}
  {arXiv preprint arXiv:1508.02595}\ } (\bibinfo {year} {2015})}\BibitemShut
  {NoStop}%
\bibitem [{\citenamefont {Lieb}\ \emph {et~al.}(1961)\citenamefont {Lieb},
  \citenamefont {Schultz},\ and\ \citenamefont {Mattis}}]{Lieb:1961aa}%
  \BibitemOpen
  \bibfield  {author} {\bibinfo {author} {\bibfnamefont {E.}~\bibnamefont
  {Lieb}}, \bibinfo {author} {\bibfnamefont {T.}~\bibnamefont {Schultz}}, \
  and\ \bibinfo {author} {\bibfnamefont {D.}~\bibnamefont {Mattis}},\ }\href
  {https://doi.org/10.1016/0003-4916(61)90115-4} {\bibfield  {journal}
  {\bibinfo  {journal} {Ann. Phys.}\ }\textbf {\bibinfo {volume} {16}},\
  \bibinfo {pages} {407} (\bibinfo {year} {1961})}\BibitemShut {NoStop}%
\bibitem [{\citenamefont {Affleck}\ and\ \citenamefont
  {Lieb}(1986)}]{Affleck:1986aa}%
  \BibitemOpen
  \bibfield  {author} {\bibinfo {author} {\bibfnamefont {I.}~\bibnamefont
  {Affleck}}\ and\ \bibinfo {author} {\bibfnamefont {E.~H.}\ \bibnamefont
  {Lieb}},\ }\href {https://doi.org/10.1007/BF00400304} {\bibfield  {journal}
  {\bibinfo  {journal} {Lett. Math. Phys.}\ }\textbf {\bibinfo {volume} {12}},\
  \bibinfo {pages} {57} (\bibinfo {year} {1986})}\BibitemShut {NoStop}%
\bibitem [{\citenamefont {Oshikawa}\ \emph {et~al.}(1997)\citenamefont
  {Oshikawa}, \citenamefont {Yamanaka},\ and\ \citenamefont
  {Affleck}}]{OYA1997}%
  \BibitemOpen
  \bibfield  {author} {\bibinfo {author} {\bibfnamefont {M.}~\bibnamefont
  {Oshikawa}}, \bibinfo {author} {\bibfnamefont {M.}~\bibnamefont {Yamanaka}},
  \ and\ \bibinfo {author} {\bibfnamefont {I.}~\bibnamefont {Affleck}},\ }\href
  {\doibase 10.1103/PhysRevLett.78.1984} {\bibfield  {journal} {\bibinfo
  {journal} {Phys. Rev. Lett.}\ }\textbf {\bibinfo {volume} {78}},\ \bibinfo
  {pages} {1984} (\bibinfo {year} {1997})}\BibitemShut {NoStop}%
\bibitem [{\citenamefont {Oshikawa}(2000)}]{Oshikawa:2000aa}%
  \BibitemOpen
  \bibfield  {author} {\bibinfo {author} {\bibfnamefont {M.}~\bibnamefont
  {Oshikawa}},\ }\href {https://link.aps.org/doi/10.1103/PhysRevLett.84.1535}
  {\bibfield  {journal} {\bibinfo  {journal} {Phys. Rev. Lett.}\ }\textbf
  {\bibinfo {volume} {84}},\ \bibinfo {pages} {1535} (\bibinfo {year}
  {2000})}\BibitemShut {NoStop}%
\bibitem [{\citenamefont {Hastings}(2004)}]{Hastings:2004ab}%
  \BibitemOpen
  \bibfield  {author} {\bibinfo {author} {\bibfnamefont {M.~B.}\ \bibnamefont
  {Hastings}},\ }\href {https://link.aps.org/doi/10.1103/PhysRevB.69.104431}
  {\bibfield  {journal} {\bibinfo  {journal} {Phys. Rev. B}\ }\textbf {\bibinfo
  {volume} {69}},\ \bibinfo {pages} {104431} (\bibinfo {year}
  {2004})}\BibitemShut {NoStop}%
\bibitem [{\citenamefont {Chen}\ \emph
  {et~al.}(2011{\natexlab{b}})\citenamefont {Chen}, \citenamefont {Gu},\ and\
  \citenamefont {Wen}}]{Chen-Gu-Wen_classification2010}%
  \BibitemOpen
  \bibfield  {author} {\bibinfo {author} {\bibfnamefont {X.}~\bibnamefont
  {Chen}}, \bibinfo {author} {\bibfnamefont {Z.-C.}\ \bibnamefont {Gu}}, \ and\
  \bibinfo {author} {\bibfnamefont {X.-G.}\ \bibnamefont {Wen}},\ }\href
  {\doibase 10.1103/PhysRevB.83.035107} {\bibfield  {journal} {\bibinfo
  {journal} {Phys. Rev. B}\ }\textbf {\bibinfo {volume} {83}},\ \bibinfo
  {pages} {035107} (\bibinfo {year} {2011}{\natexlab{b}})}\BibitemShut
  {NoStop}%
\bibitem [{\citenamefont {Fuji}(2016)}]{Fuji-SymmetryProtection-PRB2016}%
  \BibitemOpen
  \bibfield  {author} {\bibinfo {author} {\bibfnamefont {Y.}~\bibnamefont
  {Fuji}},\ }\href {\doibase 10.1103/PhysRevB.93.104425} {\bibfield  {journal}
  {\bibinfo  {journal} {Phys. Rev. B}\ }\textbf {\bibinfo {volume} {93}},\
  \bibinfo {pages} {104425} (\bibinfo {year} {2016})}\BibitemShut {NoStop}%
\bibitem [{\citenamefont {Watanabe}\ \emph {et~al.}(2015)\citenamefont
  {Watanabe}, \citenamefont {Po}, \citenamefont {Vishwanath},\ and\
  \citenamefont {Zaletel}}]{Watanabe:2015aa}%
  \BibitemOpen
  \bibfield  {author} {\bibinfo {author} {\bibfnamefont {H.}~\bibnamefont
  {Watanabe}}, \bibinfo {author} {\bibfnamefont {H.~C.}\ \bibnamefont {Po}},
  \bibinfo {author} {\bibfnamefont {A.}~\bibnamefont {Vishwanath}}, \ and\
  \bibinfo {author} {\bibfnamefont {M.}~\bibnamefont {Zaletel}},\ }\href
  {https://doi.org/10.1073/pnas.1514665112} {\bibfield  {journal} {\bibinfo
  {journal} {Proc. Natl. Acad. Sci. USA}\ }\textbf {\bibinfo {volume} {112}},\
  \bibinfo {pages} {14551} (\bibinfo {year} {2015})}\BibitemShut {NoStop}%
\bibitem [{\citenamefont {Ogata}\ and\ \citenamefont
  {Tasaki}(2019)}]{Ogata:2018aa}%
  \BibitemOpen
  \bibfield  {author} {\bibinfo {author} {\bibfnamefont {Y.}~\bibnamefont
  {Ogata}}\ and\ \bibinfo {author} {\bibfnamefont {H.}~\bibnamefont {Tasaki}},\
  }\href {\doibase 10.1007/s00220-019-03343-5} {\bibfield  {journal} {\bibinfo
  {journal} {Commun. Math. Phys.}\ }\textbf {\bibinfo {volume} {372}},\
  \bibinfo {pages} {951} (\bibinfo {year} {2019})}\BibitemShut {NoStop}%
\bibitem [{\citenamefont {Ogata}\ \emph {et~al.}(2021)\citenamefont {Ogata},
  \citenamefont {Tachikawa},\ and\ \citenamefont {Tasaki}}]{Ogata:2020aa}%
  \BibitemOpen
  \bibfield  {author} {\bibinfo {author} {\bibfnamefont {Y.}~\bibnamefont
  {Ogata}}, \bibinfo {author} {\bibfnamefont {Y.}~\bibnamefont {Tachikawa}}, \
  and\ \bibinfo {author} {\bibfnamefont {H.}~\bibnamefont {Tasaki}},\ }\href
  {\doibase 10.1007/s00220-021-04116-9} {\bibfield  {journal} {\bibinfo
  {journal} {Commun. Math. Phys.}\ }\textbf {\bibinfo {volume} {385}},\
  \bibinfo {pages} {79} (\bibinfo {year} {2021})}\BibitemShut {NoStop}%
\bibitem [{\citenamefont {Yao}\ and\ \citenamefont
  {Oshikawa}(2021)}]{Yao:2021aa}%
  \BibitemOpen
  \bibfield  {author} {\bibinfo {author} {\bibfnamefont {Y.}~\bibnamefont
  {Yao}}\ and\ \bibinfo {author} {\bibfnamefont {M.}~\bibnamefont {Oshikawa}},\
  }\href {\doibase 10.1103/PhysRevLett.126.217201} {\bibfield  {journal}
  {\bibinfo  {journal} {Phys. Rev. Lett.}\ }\textbf {\bibinfo {volume} {126}},\
  \bibinfo {pages} {217201} (\bibinfo {year} {2021})}\BibitemShut {NoStop}%
\bibitem [{\citenamefont {Yao}\ \emph {et~al.}(2019)\citenamefont {Yao},
  \citenamefont {Hsieh},\ and\ \citenamefont {Oshikawa}}]{Yao:2019aa}%
  \BibitemOpen
  \bibfield  {author} {\bibinfo {author} {\bibfnamefont {Y.}~\bibnamefont
  {Yao}}, \bibinfo {author} {\bibfnamefont {C.-T.}\ \bibnamefont {Hsieh}}, \
  and\ \bibinfo {author} {\bibfnamefont {M.}~\bibnamefont {Oshikawa}},\ }\href
  {https://doi.org/10.1103/PhysRevLett.123.180201} {\bibfield  {journal}
  {\bibinfo  {journal} {Phys. Rev. Lett.}\ }\textbf {\bibinfo {volume} {123}},\
  \bibinfo {pages} {180201} (\bibinfo {year} {2019})}\BibitemShut {NoStop}%
\bibitem [{\citenamefont {Li}\ \emph {et~al.}(2024)\citenamefont {Li},
  \citenamefont {Hsieh}, \citenamefont {Yao},\ and\ \citenamefont
  {Oshikawa}}]{PhysRevB.110.045118}%
  \BibitemOpen
  \bibfield  {author} {\bibinfo {author} {\bibfnamefont {L.}~\bibnamefont
  {Li}}, \bibinfo {author} {\bibfnamefont {C.-T.}\ \bibnamefont {Hsieh}},
  \bibinfo {author} {\bibfnamefont {Y.}~\bibnamefont {Yao}}, \ and\ \bibinfo
  {author} {\bibfnamefont {M.}~\bibnamefont {Oshikawa}},\ }\href {\doibase
  10.1103/PhysRevB.110.045118} {\bibfield  {journal} {\bibinfo  {journal}
  {Phys. Rev. B}\ }\textbf {\bibinfo {volume} {110}},\ \bibinfo {pages}
  {045118} (\bibinfo {year} {2024})}\BibitemShut {NoStop}%
\bibitem [{\citenamefont {Li}\ and\ \citenamefont
  {Yao}(2022)}]{PhysRevB.106.224420}%
  \BibitemOpen
  \bibfield  {author} {\bibinfo {author} {\bibfnamefont {L.}~\bibnamefont
  {Li}}\ and\ \bibinfo {author} {\bibfnamefont {Y.}~\bibnamefont {Yao}},\
  }\href {\doibase 10.1103/PhysRevB.106.224420} {\bibfield  {journal} {\bibinfo
   {journal} {Phys. Rev. B}\ }\textbf {\bibinfo {volume} {106}},\ \bibinfo
  {pages} {224420} (\bibinfo {year} {2022})}\BibitemShut {NoStop}%
\bibitem [{\citenamefont {Yao}\ \emph {et~al.}(2024)\citenamefont {Yao},
  \citenamefont {Li}, \citenamefont {Oshikawa},\ and\ \citenamefont
  {Hsieh}}]{PhysRevLett.133.136705}%
  \BibitemOpen
  \bibfield  {author} {\bibinfo {author} {\bibfnamefont {Y.}~\bibnamefont
  {Yao}}, \bibinfo {author} {\bibfnamefont {L.}~\bibnamefont {Li}}, \bibinfo
  {author} {\bibfnamefont {M.}~\bibnamefont {Oshikawa}}, \ and\ \bibinfo
  {author} {\bibfnamefont {C.-T.}\ \bibnamefont {Hsieh}},\ }\href {\doibase
  10.1103/PhysRevLett.133.136705} {\bibfield  {journal} {\bibinfo  {journal}
  {Phys. Rev. Lett.}\ }\textbf {\bibinfo {volume} {133}},\ \bibinfo {pages}
  {136705} (\bibinfo {year} {2024})}\BibitemShut {NoStop}%
\bibitem [{\citenamefont {Ma}\ \emph {et~al.}(1979)\citenamefont {Ma},
  \citenamefont {Dasgupta},\ and\ \citenamefont {Hu}}]{PhysRevLett.43.1434}%
  \BibitemOpen
  \bibfield  {author} {\bibinfo {author} {\bibfnamefont {S.-k.}\ \bibnamefont
  {Ma}}, \bibinfo {author} {\bibfnamefont {C.}~\bibnamefont {Dasgupta}}, \ and\
  \bibinfo {author} {\bibfnamefont {C.-k.}\ \bibnamefont {Hu}},\ }\href
  {\doibase 10.1103/PhysRevLett.43.1434} {\bibfield  {journal} {\bibinfo
  {journal} {Phys. Rev. Lett.}\ }\textbf {\bibinfo {volume} {43}},\ \bibinfo
  {pages} {1434} (\bibinfo {year} {1979})}\BibitemShut {NoStop}%
\bibitem [{\citenamefont {Dasgupta}\ and\ \citenamefont
  {Ma}(1980)}]{PhysRevB.22.1305}%
  \BibitemOpen
  \bibfield  {author} {\bibinfo {author} {\bibfnamefont {C.}~\bibnamefont
  {Dasgupta}}\ and\ \bibinfo {author} {\bibfnamefont {S.-k.}\ \bibnamefont
  {Ma}},\ }\href {\doibase 10.1103/PhysRevB.22.1305} {\bibfield  {journal}
  {\bibinfo  {journal} {Phys. Rev. B}\ }\textbf {\bibinfo {volume} {22}},\
  \bibinfo {pages} {1305} (\bibinfo {year} {1980})}\BibitemShut {NoStop}%
\bibitem [{\citenamefont {Bhatt}\ and\ \citenamefont
  {Lee}(1982)}]{PhysRevLett.48.344}%
  \BibitemOpen
  \bibfield  {author} {\bibinfo {author} {\bibfnamefont {R.~N.}\ \bibnamefont
  {Bhatt}}\ and\ \bibinfo {author} {\bibfnamefont {P.~A.}\ \bibnamefont
  {Lee}},\ }\href {\doibase 10.1103/PhysRevLett.48.344} {\bibfield  {journal}
  {\bibinfo  {journal} {Phys. Rev. Lett.}\ }\textbf {\bibinfo {volume} {48}},\
  \bibinfo {pages} {344} (\bibinfo {year} {1982})}\BibitemShut {NoStop}%
\bibitem [{\citenamefont {Sandvik}\ and\ \citenamefont
  {Veki\ifmmode~\acute{c}\else \'{c}\fi{}}(1995)}]{PhysRevLett.74.1226}%
  \BibitemOpen
  \bibfield  {author} {\bibinfo {author} {\bibfnamefont {A.~W.}\ \bibnamefont
  {Sandvik}}\ and\ \bibinfo {author} {\bibfnamefont {M.}~\bibnamefont
  {Veki\ifmmode~\acute{c}\else \'{c}\fi{}}},\ }\href {\doibase
  10.1103/PhysRevLett.74.1226} {\bibfield  {journal} {\bibinfo  {journal}
  {Phys. Rev. Lett.}\ }\textbf {\bibinfo {volume} {74}},\ \bibinfo {pages}
  {1226} (\bibinfo {year} {1995})}\BibitemShut {NoStop}%
\bibitem [{\citenamefont {Bobroff}\ \emph {et~al.}(2009)\citenamefont
  {Bobroff}, \citenamefont {Laflorencie}, \citenamefont {Alexander},
  \citenamefont {Mahajan}, \citenamefont {Koteswararao},\ and\ \citenamefont
  {Mendels}}]{PhysRevLett.103.047201}%
  \BibitemOpen
  \bibfield  {author} {\bibinfo {author} {\bibfnamefont {J.}~\bibnamefont
  {Bobroff}}, \bibinfo {author} {\bibfnamefont {N.}~\bibnamefont
  {Laflorencie}}, \bibinfo {author} {\bibfnamefont {L.~K.}\ \bibnamefont
  {Alexander}}, \bibinfo {author} {\bibfnamefont {A.~V.}\ \bibnamefont
  {Mahajan}}, \bibinfo {author} {\bibfnamefont {B.}~\bibnamefont
  {Koteswararao}}, \ and\ \bibinfo {author} {\bibfnamefont {P.}~\bibnamefont
  {Mendels}},\ }\href {\doibase 10.1103/PhysRevLett.103.047201} {\bibfield
  {journal} {\bibinfo  {journal} {Phys. Rev. Lett.}\ }\textbf {\bibinfo
  {volume} {103}},\ \bibinfo {pages} {047201} (\bibinfo {year}
  {2009})}\BibitemShut {NoStop}%
\bibitem [{\citenamefont {Preskill}(2018)}]{preskill2018quantum}%
  \BibitemOpen
  \bibfield  {author} {\bibinfo {author} {\bibfnamefont {J.}~\bibnamefont
  {Preskill}},\ }\href {https://doi.org/10.22331/q-2018-08-06-79} {\bibfield
  {journal} {\bibinfo  {journal} {Quantum}\ }\textbf {\bibinfo {volume} {2}},\
  \bibinfo {pages} {79} (\bibinfo {year} {2018})}\BibitemShut {NoStop}%
\bibitem [{\citenamefont {Cattaneo}\ \emph {et~al.}(2021)\citenamefont
  {Cattaneo}, \citenamefont {De~Chiara}, \citenamefont {Maniscalco},
  \citenamefont {Zambrini},\ and\ \citenamefont
  {Giorgi}}]{PhysRevLett.126.130403}%
  \BibitemOpen
  \bibfield  {author} {\bibinfo {author} {\bibfnamefont {M.}~\bibnamefont
  {Cattaneo}}, \bibinfo {author} {\bibfnamefont {G.}~\bibnamefont {De~Chiara}},
  \bibinfo {author} {\bibfnamefont {S.}~\bibnamefont {Maniscalco}}, \bibinfo
  {author} {\bibfnamefont {R.}~\bibnamefont {Zambrini}}, \ and\ \bibinfo
  {author} {\bibfnamefont {G.~L.}\ \bibnamefont {Giorgi}},\ }\href {\doibase
  10.1103/PhysRevLett.126.130403} {\bibfield  {journal} {\bibinfo  {journal}
  {Phys. Rev. Lett.}\ }\textbf {\bibinfo {volume} {126}},\ \bibinfo {pages}
  {130403} (\bibinfo {year} {2021})}\BibitemShut {NoStop}%
\bibitem [{\citenamefont {Schlimgen}\ \emph {et~al.}(2021)\citenamefont
  {Schlimgen}, \citenamefont {Head-Marsden}, \citenamefont {Sager},
  \citenamefont {Narang},\ and\ \citenamefont
  {Mazziotti}}]{PhysRevLett.127.270503}%
  \BibitemOpen
  \bibfield  {author} {\bibinfo {author} {\bibfnamefont {A.~W.}\ \bibnamefont
  {Schlimgen}}, \bibinfo {author} {\bibfnamefont {K.}~\bibnamefont
  {Head-Marsden}}, \bibinfo {author} {\bibfnamefont {L.~M.}\ \bibnamefont
  {Sager}}, \bibinfo {author} {\bibfnamefont {P.}~\bibnamefont {Narang}}, \
  and\ \bibinfo {author} {\bibfnamefont {D.~A.}\ \bibnamefont {Mazziotti}},\
  }\href {\doibase 10.1103/PhysRevLett.127.270503} {\bibfield  {journal}
  {\bibinfo  {journal} {Phys. Rev. Lett.}\ }\textbf {\bibinfo {volume} {127}},\
  \bibinfo {pages} {270503} (\bibinfo {year} {2021})}\BibitemShut {NoStop}%
\bibitem [{\citenamefont {Guo}\ and\ \citenamefont
  {Yang}(2022)}]{PRXQuantum.3.040313}%
  \BibitemOpen
  \bibfield  {author} {\bibinfo {author} {\bibfnamefont {Y.}~\bibnamefont
  {Guo}}\ and\ \bibinfo {author} {\bibfnamefont {S.}~\bibnamefont {Yang}},\
  }\href {\doibase 10.1103/PRXQuantum.3.040313} {\bibfield  {journal} {\bibinfo
   {journal} {PRX Quantum}\ }\textbf {\bibinfo {volume} {3}},\ \bibinfo {pages}
  {040313} (\bibinfo {year} {2022})}\BibitemShut {NoStop}%
\bibitem [{\citenamefont {Sang}\ \emph {et~al.}(2024)\citenamefont {Sang},
  \citenamefont {Zou},\ and\ \citenamefont {Hsieh}}]{PhysRevX.14.031044}%
  \BibitemOpen
  \bibfield  {author} {\bibinfo {author} {\bibfnamefont {S.}~\bibnamefont
  {Sang}}, \bibinfo {author} {\bibfnamefont {Y.}~\bibnamefont {Zou}}, \ and\
  \bibinfo {author} {\bibfnamefont {T.~H.}\ \bibnamefont {Hsieh}},\ }\href
  {\doibase 10.1103/PhysRevX.14.031044} {\bibfield  {journal} {\bibinfo
  {journal} {Phys. Rev. X}\ }\textbf {\bibinfo {volume} {14}},\ \bibinfo
  {pages} {031044} (\bibinfo {year} {2024})}\BibitemShut {NoStop}%
\bibitem [{\citenamefont {Sang}\ and\ \citenamefont
  {Hsieh}(2025)}]{PhysRevLett.134.070403}%
  \BibitemOpen
  \bibfield  {author} {\bibinfo {author} {\bibfnamefont {S.}~\bibnamefont
  {Sang}}\ and\ \bibinfo {author} {\bibfnamefont {T.~H.}\ \bibnamefont
  {Hsieh}},\ }\href {\doibase 10.1103/PhysRevLett.134.070403} {\bibfield
  {journal} {\bibinfo  {journal} {Phys. Rev. Lett.}\ }\textbf {\bibinfo
  {volume} {134}},\ \bibinfo {pages} {070403} (\bibinfo {year}
  {2025})}\BibitemShut {NoStop}%
\bibitem [{\citenamefont {Yang}\ \emph {et~al.}(2025)\citenamefont {Yang},
  \citenamefont {Shi},\ and\ \citenamefont {Lee}}]{yang2025topological}%
  \BibitemOpen
  \bibfield  {author} {\bibinfo {author} {\bibfnamefont {T.-H.}\ \bibnamefont
  {Yang}}, \bibinfo {author} {\bibfnamefont {B.}~\bibnamefont {Shi}}, \ and\
  \bibinfo {author} {\bibfnamefont {J.~Y.}\ \bibnamefont {Lee}},\ }\href
  {https://arxiv.org/abs/2506.04221} {\bibfield  {journal} {\bibinfo  {journal}
  {arXiv preprint arXiv:2506.04221}\ } (\bibinfo {year} {2025})}\BibitemShut
  {NoStop}%
\bibitem [{\citenamefont {Sang}\ \emph {et~al.}(2025)\citenamefont {Sang},
  \citenamefont {Lessa}, \citenamefont {Mong}, \citenamefont {Grover},
  \citenamefont {Wang},\ and\ \citenamefont {Hsieh}}]{sang2025mixed}%
  \BibitemOpen
  \bibfield  {author} {\bibinfo {author} {\bibfnamefont {S.}~\bibnamefont
  {Sang}}, \bibinfo {author} {\bibfnamefont {L.~A.}\ \bibnamefont {Lessa}},
  \bibinfo {author} {\bibfnamefont {R.~S.}\ \bibnamefont {Mong}}, \bibinfo
  {author} {\bibfnamefont {T.}~\bibnamefont {Grover}}, \bibinfo {author}
  {\bibfnamefont {C.}~\bibnamefont {Wang}}, \ and\ \bibinfo {author}
  {\bibfnamefont {T.~H.}\ \bibnamefont {Hsieh}},\ }\href
  {https://arxiv.org/abs/2507.02292} {\bibfield  {journal} {\bibinfo  {journal}
  {arXiv preprint arXiv:2507.02292}\ } (\bibinfo {year} {2025})}\BibitemShut
  {NoStop}%
\bibitem [{\citenamefont {Lessa}\ \emph
  {et~al.}(2025{\natexlab{a}})\citenamefont {Lessa}, \citenamefont {Ma},
  \citenamefont {Zhang}, \citenamefont {Bi}, \citenamefont {Cheng},\ and\
  \citenamefont {Wang}}]{PRXQuantum.6.010344}%
  \BibitemOpen
  \bibfield  {author} {\bibinfo {author} {\bibfnamefont {L.~A.}\ \bibnamefont
  {Lessa}}, \bibinfo {author} {\bibfnamefont {R.}~\bibnamefont {Ma}}, \bibinfo
  {author} {\bibfnamefont {J.-H.}\ \bibnamefont {Zhang}}, \bibinfo {author}
  {\bibfnamefont {Z.}~\bibnamefont {Bi}}, \bibinfo {author} {\bibfnamefont
  {M.}~\bibnamefont {Cheng}}, \ and\ \bibinfo {author} {\bibfnamefont
  {C.}~\bibnamefont {Wang}},\ }\href {\doibase 10.1103/PRXQuantum.6.010344}
  {\bibfield  {journal} {\bibinfo  {journal} {PRX Quantum}\ }\textbf {\bibinfo
  {volume} {6}},\ \bibinfo {pages} {010344} (\bibinfo {year}
  {2025}{\natexlab{a}})}\BibitemShut {NoStop}%
\bibitem [{\citenamefont {Gu}\ \emph {et~al.}(2024)\citenamefont {Gu},
  \citenamefont {Wang},\ and\ \citenamefont {Wang}}]{gu2024spontaneous}%
  \BibitemOpen
  \bibfield  {author} {\bibinfo {author} {\bibfnamefont {D.}~\bibnamefont
  {Gu}}, \bibinfo {author} {\bibfnamefont {Z.}~\bibnamefont {Wang}}, \ and\
  \bibinfo {author} {\bibfnamefont {Z.}~\bibnamefont {Wang}},\ }\href
  {https://arxiv.org/abs/2406.19381} {\bibfield  {journal} {\bibinfo  {journal}
  {arXiv preprint arXiv:2406.19381}\ } (\bibinfo {year} {2024})}\BibitemShut
  {NoStop}%
\bibitem [{\citenamefont {Wang}\ \emph {et~al.}(2025)\citenamefont {Wang},
  \citenamefont {Wu},\ and\ \citenamefont {Wang}}]{PRXQuantum.6.010314}%
  \BibitemOpen
  \bibfield  {author} {\bibinfo {author} {\bibfnamefont {Z.}~\bibnamefont
  {Wang}}, \bibinfo {author} {\bibfnamefont {Z.}~\bibnamefont {Wu}}, \ and\
  \bibinfo {author} {\bibfnamefont {Z.}~\bibnamefont {Wang}},\ }\href {\doibase
  10.1103/PRXQuantum.6.010314} {\bibfield  {journal} {\bibinfo  {journal} {PRX
  Quantum}\ }\textbf {\bibinfo {volume} {6}},\ \bibinfo {pages} {010314}
  (\bibinfo {year} {2025})}\BibitemShut {NoStop}%
\bibitem [{\citenamefont {Ellison}\ and\ \citenamefont
  {Cheng}(2025)}]{PRXQuantum.6.010315}%
  \BibitemOpen
  \bibfield  {author} {\bibinfo {author} {\bibfnamefont {T.~D.}\ \bibnamefont
  {Ellison}}\ and\ \bibinfo {author} {\bibfnamefont {M.}~\bibnamefont
  {Cheng}},\ }\href {\doibase 10.1103/PRXQuantum.6.010315} {\bibfield
  {journal} {\bibinfo  {journal} {PRX Quantum}\ }\textbf {\bibinfo {volume}
  {6}},\ \bibinfo {pages} {010315} (\bibinfo {year} {2025})}\BibitemShut
  {NoStop}%
\bibitem [{\citenamefont {Sohal}\ and\ \citenamefont
  {Prem}(2025)}]{PRXQuantum.6.010313}%
  \BibitemOpen
  \bibfield  {author} {\bibinfo {author} {\bibfnamefont {R.}~\bibnamefont
  {Sohal}}\ and\ \bibinfo {author} {\bibfnamefont {A.}~\bibnamefont {Prem}},\
  }\href {\doibase 10.1103/PRXQuantum.6.010313} {\bibfield  {journal} {\bibinfo
   {journal} {PRX Quantum}\ }\textbf {\bibinfo {volume} {6}},\ \bibinfo {pages}
  {010313} (\bibinfo {year} {2025})}\BibitemShut {NoStop}%
\bibitem [{\citenamefont {Luo}\ \emph {et~al.}(2025)\citenamefont {Luo},
  \citenamefont {Wang},\ and\ \citenamefont {Bi}}]{luo2025topological}%
  \BibitemOpen
  \bibfield  {author} {\bibinfo {author} {\bibfnamefont {R.}~\bibnamefont
  {Luo}}, \bibinfo {author} {\bibfnamefont {Y.-N.}\ \bibnamefont {Wang}}, \
  and\ \bibinfo {author} {\bibfnamefont {Z.}~\bibnamefont {Bi}},\ }\href
  {https://arxiv.org/abs/2507.06218} {\bibfield  {journal} {\bibinfo  {journal}
  {arXiv preprint arXiv:2507.06218}\ } (\bibinfo {year} {2025})}\BibitemShut
  {NoStop}%
\bibitem [{\citenamefont {Lee}\ \emph {et~al.}(2025)\citenamefont {Lee},
  \citenamefont {You},\ and\ \citenamefont {Xu}}]{Lee_2025}%
  \BibitemOpen
  \bibfield  {author} {\bibinfo {author} {\bibfnamefont {J.~Y.}\ \bibnamefont
  {Lee}}, \bibinfo {author} {\bibfnamefont {Y.-Z.}\ \bibnamefont {You}}, \ and\
  \bibinfo {author} {\bibfnamefont {C.}~\bibnamefont {Xu}},\ }\href {\doibase
  10.22331/q-2025-01-23-1607} {\bibfield  {journal} {\bibinfo  {journal}
  {Quantum}\ }\textbf {\bibinfo {volume} {9}},\ \bibinfo {pages} {1607}
  (\bibinfo {year} {2025})}\BibitemShut {NoStop}%
\bibitem [{\citenamefont {de~Groot}\ \emph {et~al.}(2022)\citenamefont
  {de~Groot}, \citenamefont {Turzillo},\ and\ \citenamefont
  {Schuch}}]{de2022symmetry}%
  \BibitemOpen
  \bibfield  {author} {\bibinfo {author} {\bibfnamefont {C.}~\bibnamefont
  {de~Groot}}, \bibinfo {author} {\bibfnamefont {A.}~\bibnamefont {Turzillo}},
  \ and\ \bibinfo {author} {\bibfnamefont {N.}~\bibnamefont {Schuch}},\ }\href
  {https://doi.org/10.22331/q-2022-11-10-856} {\bibfield  {journal} {\bibinfo
  {journal} {Quantum}\ }\textbf {\bibinfo {volume} {6}},\ \bibinfo {pages}
  {856} (\bibinfo {year} {2022})}\BibitemShut {NoStop}%
\bibitem [{\citenamefont {Zhang}\ \emph {et~al.}(2022)\citenamefont {Zhang},
  \citenamefont {Qi},\ and\ \citenamefont {Bi}}]{zhang2022strange}%
  \BibitemOpen
  \bibfield  {author} {\bibinfo {author} {\bibfnamefont {J.-H.}\ \bibnamefont
  {Zhang}}, \bibinfo {author} {\bibfnamefont {Y.}~\bibnamefont {Qi}}, \ and\
  \bibinfo {author} {\bibfnamefont {Z.}~\bibnamefont {Bi}},\ }\href
  {https://arxiv.org/abs/2210.17485} {\bibfield  {journal} {\bibinfo  {journal}
  {arXiv preprint arXiv:2210.17485}\ } (\bibinfo {year} {2022})}\BibitemShut
  {NoStop}%
\bibitem [{\citenamefont {Ma}\ and\ \citenamefont
  {Wang}(2023)}]{ma2023average}%
  \BibitemOpen
  \bibfield  {author} {\bibinfo {author} {\bibfnamefont {R.}~\bibnamefont
  {Ma}}\ and\ \bibinfo {author} {\bibfnamefont {C.}~\bibnamefont {Wang}},\
  }\href {https://doi.org/10.1103/PhysRevX.13.031016} {\bibfield  {journal}
  {\bibinfo  {journal} {Phys. Rev. X}\ }\textbf {\bibinfo {volume} {13}},\
  \bibinfo {pages} {031016} (\bibinfo {year} {2023})}\BibitemShut {NoStop}%
\bibitem [{\citenamefont {Ma}\ \emph {et~al.}(2025)\citenamefont {Ma},
  \citenamefont {Zhang}, \citenamefont {Bi}, \citenamefont {Cheng},\ and\
  \citenamefont {Wang}}]{ma2025topological}%
  \BibitemOpen
  \bibfield  {author} {\bibinfo {author} {\bibfnamefont {R.}~\bibnamefont
  {Ma}}, \bibinfo {author} {\bibfnamefont {J.-H.}\ \bibnamefont {Zhang}},
  \bibinfo {author} {\bibfnamefont {Z.}~\bibnamefont {Bi}}, \bibinfo {author}
  {\bibfnamefont {M.}~\bibnamefont {Cheng}}, \ and\ \bibinfo {author}
  {\bibfnamefont {C.}~\bibnamefont {Wang}},\ }\href
  {https://doi.org/10.1103/PhysRevX.15.021062} {\bibfield  {journal} {\bibinfo
  {journal} {Phys. Rev. X}\ }\textbf {\bibinfo {volume} {15}},\ \bibinfo
  {pages} {021062} (\bibinfo {year} {2025})}\BibitemShut {NoStop}%
\bibitem [{\citenamefont {Guo}\ \emph {et~al.}(2025)\citenamefont {Guo},
  \citenamefont {Zhang}, \citenamefont {Zhang}, \citenamefont {Yang},\ and\
  \citenamefont {Bi}}]{PhysRevX.15.021060}%
  \BibitemOpen
  \bibfield  {author} {\bibinfo {author} {\bibfnamefont {Y.}~\bibnamefont
  {Guo}}, \bibinfo {author} {\bibfnamefont {J.-H.}\ \bibnamefont {Zhang}},
  \bibinfo {author} {\bibfnamefont {H.-R.}\ \bibnamefont {Zhang}}, \bibinfo
  {author} {\bibfnamefont {S.}~\bibnamefont {Yang}}, \ and\ \bibinfo {author}
  {\bibfnamefont {Z.}~\bibnamefont {Bi}},\ }\href {\doibase
  10.1103/PhysRevX.15.021060} {\bibfield  {journal} {\bibinfo  {journal} {Phys.
  Rev. X}\ }\textbf {\bibinfo {volume} {15}},\ \bibinfo {pages} {021060}
  (\bibinfo {year} {2025})}\BibitemShut {NoStop}%
\bibitem [{\citenamefont {Xue}\ \emph {et~al.}(2024)\citenamefont {Xue},
  \citenamefont {Lee},\ and\ \citenamefont {Bao}}]{xue2024tensor}%
  \BibitemOpen
  \bibfield  {author} {\bibinfo {author} {\bibfnamefont {H.}~\bibnamefont
  {Xue}}, \bibinfo {author} {\bibfnamefont {J.~Y.}\ \bibnamefont {Lee}}, \ and\
  \bibinfo {author} {\bibfnamefont {Y.}~\bibnamefont {Bao}},\ }\href
  {https://arxiv.org/abs/2403.17069} {\bibfield  {journal} {\bibinfo  {journal}
  {arXiv preprint arXiv:2403.17069}\ } (\bibinfo {year} {2024})}\BibitemShut
  {NoStop}%
\bibitem [{\citenamefont {You}\ and\ \citenamefont
  {Oshikawa}(2024)}]{you2024intrinsic}%
  \BibitemOpen
  \bibfield  {author} {\bibinfo {author} {\bibfnamefont {Y.}~\bibnamefont
  {You}}\ and\ \bibinfo {author} {\bibfnamefont {M.}~\bibnamefont {Oshikawa}},\
  }\href {https://arxiv.org/abs/2407.08786} {\bibfield  {journal} {\bibinfo
  {journal} {arXiv preprint arXiv:2407.08786}\ } (\bibinfo {year}
  {2024})}\BibitemShut {NoStop}%
\bibitem [{\citenamefont {Sun}\ \emph {et~al.}(2025)\citenamefont {Sun},
  \citenamefont {Zhang}, \citenamefont {Bi},\ and\ \citenamefont
  {You}}]{PRXQuantum.6.020333}%
  \BibitemOpen
  \bibfield  {author} {\bibinfo {author} {\bibfnamefont {S.}~\bibnamefont
  {Sun}}, \bibinfo {author} {\bibfnamefont {J.-H.}\ \bibnamefont {Zhang}},
  \bibinfo {author} {\bibfnamefont {Z.}~\bibnamefont {Bi}}, \ and\ \bibinfo
  {author} {\bibfnamefont {Y.}~\bibnamefont {You}},\ }\href {\doibase
  10.1103/PRXQuantum.6.020333} {\bibfield  {journal} {\bibinfo  {journal} {PRX
  Quantum}\ }\textbf {\bibinfo {volume} {6}},\ \bibinfo {pages} {020333}
  (\bibinfo {year} {2025})}\BibitemShut {NoStop}%
\bibitem [{\citenamefont {Kawabata}\ \emph {et~al.}(2024)\citenamefont
  {Kawabata}, \citenamefont {Sohal},\ and\ \citenamefont
  {Ryu}}]{PhysRevLett.132.070402}%
  \BibitemOpen
  \bibfield  {author} {\bibinfo {author} {\bibfnamefont {K.}~\bibnamefont
  {Kawabata}}, \bibinfo {author} {\bibfnamefont {R.}~\bibnamefont {Sohal}}, \
  and\ \bibinfo {author} {\bibfnamefont {S.}~\bibnamefont {Ryu}},\ }\href
  {\doibase 10.1103/PhysRevLett.132.070402} {\bibfield  {journal} {\bibinfo
  {journal} {Phys. Rev. Lett.}\ }\textbf {\bibinfo {volume} {132}},\ \bibinfo
  {pages} {070402} (\bibinfo {year} {2024})}\BibitemShut {NoStop}%
\bibitem [{\citenamefont {Lessa}\ \emph
  {et~al.}(2025{\natexlab{b}})\citenamefont {Lessa}, \citenamefont {Cheng},\
  and\ \citenamefont {Wang}}]{PhysRevX.15.011069}%
  \BibitemOpen
  \bibfield  {author} {\bibinfo {author} {\bibfnamefont {L.~A.}\ \bibnamefont
  {Lessa}}, \bibinfo {author} {\bibfnamefont {M.}~\bibnamefont {Cheng}}, \ and\
  \bibinfo {author} {\bibfnamefont {C.}~\bibnamefont {Wang}},\ }\href {\doibase
  10.1103/PhysRevX.15.011069} {\bibfield  {journal} {\bibinfo  {journal} {Phys.
  Rev. X}\ }\textbf {\bibinfo {volume} {15}},\ \bibinfo {pages} {011069}
  (\bibinfo {year} {2025}{\natexlab{b}})}\BibitemShut {NoStop}%
\bibitem [{\citenamefont {Zang}\ \emph {et~al.}(2024)\citenamefont {Zang},
  \citenamefont {Gu},\ and\ \citenamefont {Jiang}}]{PhysRevLett.133.106503}%
  \BibitemOpen
  \bibfield  {author} {\bibinfo {author} {\bibfnamefont {Y.}~\bibnamefont
  {Zang}}, \bibinfo {author} {\bibfnamefont {Y.}~\bibnamefont {Gu}}, \ and\
  \bibinfo {author} {\bibfnamefont {S.}~\bibnamefont {Jiang}},\ }\href
  {\doibase 10.1103/PhysRevLett.133.106503} {\bibfield  {journal} {\bibinfo
  {journal} {Phys. Rev. Lett.}\ }\textbf {\bibinfo {volume} {133}},\ \bibinfo
  {pages} {106503} (\bibinfo {year} {2024})}\BibitemShut {NoStop}%
\bibitem [{\citenamefont {Wang}\ and\ \citenamefont
  {Li}(2025)}]{PRXQuantum.6.010347}%
  \BibitemOpen
  \bibfield  {author} {\bibinfo {author} {\bibfnamefont {Z.}~\bibnamefont
  {Wang}}\ and\ \bibinfo {author} {\bibfnamefont {L.}~\bibnamefont {Li}},\
  }\href {\doibase 10.1103/PRXQuantum.6.010347} {\bibfield  {journal} {\bibinfo
   {journal} {PRX Quantum}\ }\textbf {\bibinfo {volume} {6}},\ \bibinfo {pages}
  {010347} (\bibinfo {year} {2025})}\BibitemShut {NoStop}%
\bibitem [{\citenamefont {Su}\ \emph {et~al.}(2024)\citenamefont {Su},
  \citenamefont {Yao},\ and\ \citenamefont
  {Furusaki}}]{PhysRevLett.133.266705}%
  \BibitemOpen
  \bibfield  {author} {\bibinfo {author} {\bibfnamefont {H.}~\bibnamefont
  {Su}}, \bibinfo {author} {\bibfnamefont {Y.}~\bibnamefont {Yao}}, \ and\
  \bibinfo {author} {\bibfnamefont {A.}~\bibnamefont {Furusaki}},\ }\href
  {\doibase 10.1103/PhysRevLett.133.266705} {\bibfield  {journal} {\bibinfo
  {journal} {Phys. Rev. Lett.}\ }\textbf {\bibinfo {volume} {133}},\ \bibinfo
  {pages} {266705} (\bibinfo {year} {2024})}\BibitemShut {NoStop}%
\bibitem [{\citenamefont {Verstraete}\ \emph {et~al.}(2006)\citenamefont
  {Verstraete}, \citenamefont {Wolf}, \citenamefont {Perez-Garcia},\ and\
  \citenamefont {Cirac}}]{PhysRevLett.96.220601}%
  \BibitemOpen
  \bibfield  {author} {\bibinfo {author} {\bibfnamefont {F.}~\bibnamefont
  {Verstraete}}, \bibinfo {author} {\bibfnamefont {M.~M.}\ \bibnamefont
  {Wolf}}, \bibinfo {author} {\bibfnamefont {D.}~\bibnamefont {Perez-Garcia}},
  \ and\ \bibinfo {author} {\bibfnamefont {J.~I.}\ \bibnamefont {Cirac}},\
  }\href {\doibase 10.1103/PhysRevLett.96.220601} {\bibfield  {journal}
  {\bibinfo  {journal} {Phys. Rev. Lett.}\ }\textbf {\bibinfo {volume} {96}},\
  \bibinfo {pages} {220601} (\bibinfo {year} {2006})}\BibitemShut {NoStop}%
\bibitem [{\citenamefont {Perez-Garcia}\ \emph {et~al.}(2006)\citenamefont
  {Perez-Garcia}, \citenamefont {Verstraete}, \citenamefont {Wolf},\ and\
  \citenamefont {Cirac}}]{perez2006matrix}%
  \BibitemOpen
  \bibfield  {author} {\bibinfo {author} {\bibfnamefont {D.}~\bibnamefont
  {Perez-Garcia}}, \bibinfo {author} {\bibfnamefont {F.}~\bibnamefont
  {Verstraete}}, \bibinfo {author} {\bibfnamefont {M.~M.}\ \bibnamefont
  {Wolf}}, \ and\ \bibinfo {author} {\bibfnamefont {J.~I.}\ \bibnamefont
  {Cirac}},\ }\href {https://arxiv.org/abs/quant-ph/0608197} {\bibfield
  {journal} {\bibinfo  {journal} {arXiv preprint quant-ph/0608197}\ } (\bibinfo
  {year} {2006})}\BibitemShut {NoStop}%
\bibitem [{\citenamefont {Cirac}\ \emph {et~al.}(2021)\citenamefont {Cirac},
  \citenamefont {P\'erez-Garc\'{\i}a}, \citenamefont {Schuch},\ and\
  \citenamefont {Verstraete}}]{RevModPhys.93.045003}%
  \BibitemOpen
  \bibfield  {author} {\bibinfo {author} {\bibfnamefont {J.~I.}\ \bibnamefont
  {Cirac}}, \bibinfo {author} {\bibfnamefont {D.}~\bibnamefont
  {P\'erez-Garc\'{\i}a}}, \bibinfo {author} {\bibfnamefont {N.}~\bibnamefont
  {Schuch}}, \ and\ \bibinfo {author} {\bibfnamefont {F.}~\bibnamefont
  {Verstraete}},\ }\href {\doibase 10.1103/RevModPhys.93.045003} {\bibfield
  {journal} {\bibinfo  {journal} {Rev. Mod. Phys.}\ }\textbf {\bibinfo {volume}
  {93}},\ \bibinfo {pages} {045003} (\bibinfo {year} {2021})}\BibitemShut
  {NoStop}%
\bibitem [{\citenamefont {Fannes}\ \emph {et~al.}(1992)\citenamefont {Fannes},
  \citenamefont {Nachtergaele},\ and\ \citenamefont {Werner}}]{Fannes:1990ur}%
  \BibitemOpen
  \bibfield  {author} {\bibinfo {author} {\bibfnamefont {M.}~\bibnamefont
  {Fannes}}, \bibinfo {author} {\bibfnamefont {B.}~\bibnamefont
  {Nachtergaele}}, \ and\ \bibinfo {author} {\bibfnamefont {R.~F.}\
  \bibnamefont {Werner}},\ }\href {\doibase 10.1007/BF02099178} {\bibfield
  {journal} {\bibinfo  {journal} {Commun. Math. Phys.}\ }\textbf {\bibinfo
  {volume} {144}},\ \bibinfo {pages} {443} (\bibinfo {year}
  {1992})}\BibitemShut {NoStop}%
\bibitem [{Note1()}]{Note1}%
  \BibitemOpen
  \bibinfo {note} {{Injectivity length $k$ refers to the smallest integer such
  that the tensors $A^{[n]}_{s_n}\protect \cdots A^{[n+k]}_{s_{n+k}}$ is
  injective when viewed as a linear transformation $\protect \mathbb
  {C}^D\rightarrow \protect \mathbb {C}^{d^k}\times \protect \mathbb {C}^D$,
  where $D$ and $d$ are the dimensions of virtual space and physical space,
  respectively.}}\BibitemShut {Stop}%
\bibitem [{Note2()}]{Note2}%
  \BibitemOpen
  \bibinfo {note} {This can be easily verified by performing the Schmidt
  decomposition of $|\Phi \rangle $ under the bipartition $S\cup
  AE$}\BibitemShut {NoStop}%
\bibitem [{Note3()}]{Note3}%
  \BibitemOpen
  \bibinfo {note} {It is generally not $s$-differentiable for multi-layer
  cases.}\BibitemShut {Stop}%
\bibitem [{Note4()}]{Note4}%
  \BibitemOpen
  \bibinfo {note} {We note that a refined definition of mixed-state phase
  equivalence—based on one-way connectivity via locally reversible channel
  circuits—was recently introduced in Ref. \cite {sang2025mixed}, motivated
  by the existence of a counterexample in two-dimensional classical statistical
  mechanics under the original two-way connectivity definition. It has been
  shown in \cite {PhysRevLett.134.070403} that two states being equivalent
  under this refined definition implies their equivalence under two-way
  connectivity. Thus two states with distinct values of $\protect \mathcal {I}$
  must also belong to distinct mixed-state matter of phases even under this
  refined definition.}\BibitemShut {Stop}%
\bibitem [{sup()}]{supplemental}%
  \BibitemOpen
  \href@noop {} {{\bibinfo {title} {{See Supplemental
  Materials.}}}}\BibitemShut {Stop}%
\bibitem [{Note5()}]{Note5}%
  \BibitemOpen
  \bibinfo {note} {The length $L$ must be even due to $\protect \bm
  {S}^z_\protect \text {tot}\rho =0$.}\BibitemShut {Stop}%
\bibitem [{\citenamefont {Schmoll}\ \emph {et~al.}(2019)\citenamefont
  {Schmoll}, \citenamefont {Haller}, \citenamefont {Rizzi},\ and\ \citenamefont
  {Or\'us}}]{PhysRevB.99.205121}%
  \BibitemOpen
  \bibfield  {author} {\bibinfo {author} {\bibfnamefont {P.}~\bibnamefont
  {Schmoll}}, \bibinfo {author} {\bibfnamefont {A.}~\bibnamefont {Haller}},
  \bibinfo {author} {\bibfnamefont {M.}~\bibnamefont {Rizzi}}, \ and\ \bibinfo
  {author} {\bibfnamefont {R.}~\bibnamefont {Or\'us}},\ }\href {\doibase
  10.1103/PhysRevB.99.205121} {\bibfield  {journal} {\bibinfo  {journal} {Phys.
  Rev. B}\ }\textbf {\bibinfo {volume} {99}},\ \bibinfo {pages} {205121}
  (\bibinfo {year} {2019})}\BibitemShut {NoStop}%
\end{thebibliography}

%

\newpage
\clearpage
\appendix

\widetext

\appendix
\appendix

\section{Mixed topological phases and phase transitions}
We consider a spin-1/2 chain with an even length $L=2N$ under the periodic boundary condition (PBC).
The clean Hamiltonian is the dimer model:
\begin{eqnarray}
H_0=-\sum_{n=1}^N(2P_{2n,2n+1}-1),
\end{eqnarray}
where $P_{2n,2n+1}$ is the singlet projection operator on the sites $2n$ and $2n+1$,
and $H_0$ has the unique ground state as dimer formed by $(2n,2n+1)$ sites.

We add the following disorder
\begin{eqnarray}
H'(B)=\sum_{n=1}^N2h_n(-S^z_{2n+1}+S^z_{2n+2}),
\end{eqnarray}
where $h_n$ are independent random variable with $B\geq0$:
\begin{eqnarray}
P(h_n=B)=50\%,\,\,P(h_n=-B)=50\%.
\end{eqnarray}

We define each ``cell'' $j$ as $2n+1$ and $2n+2$ lattice points and then the concept of product state in {\bf Theorem~5} can be defined.
At $B=0$, the ground state of the system can be prepared from a $\Z^x_2$ symmetric product state $|\rightarrow\leftarrow\cdots\rightarrow\leftarrow\rangle$ which is obviously a product of mixed (actually pure) states of each cell,
by the following $\Z^x_2$ symmetric FDLUC $\prod_{n}U_{2n,2n+1}$ where 
\begin{eqnarray}
U_{2n,2n+1}&=&\frac{1}{\sqrt{2}}(|\rightarrow\leftarrow\rangle-|\leftarrow\rightarrow\rangle) \langle\rightarrow\leftarrow|+\frac{1}{\sqrt{2}}( |\rightarrow\leftarrow\rangle+|\leftarrow\rightarrow\rangle)\langle\leftarrow\rightarrow|\nonumber\\
&&+|\leftarrow\leftarrow\rangle\langle \leftarrow\leftarrow|+|\rightarrow\rightarrow\rangle\langle \rightarrow\rightarrow|.
\end{eqnarray}
At $B=\infty$, the density matrix is already a product mixed state under the current definition of cells.

Let us calculate the average value of the twisting operator
\begin{eqnarray}
U=\exp\left(\frac{2\pi i}{L}\sum_{n=1}^LnS^z_n\right),
\end{eqnarray}
and a local staggering magnetism:
\begin{eqnarray}
M_s\equiv-\frac{4}{N}\sum_{n=1}^NS^z_{2n}S^z_{2n+1}.
\end{eqnarray}

\subsection{Calculation of a general local operator}
Let us calculate the average value of a general product operator $\otimes_{n=1}^NO_n$, in which $O_n$ includes the Hilbert space at the sites $2n$ and $2n+1$.

Obviously,
the Hamiltonian,
at any arbitrary realization $H_0+H'(B)$,
has a ground state as a product of states $\Psi_n$ which stays within the Hilbert space of the sites $2n$ and $2n+1$.
Focusing on this pair of sites,
we define the ``disorder'' pseudospin space
\begin{eqnarray}
(a)_{p,q}=h_{n-1}+h_n=\left(\begin{array}{cc}2B&0\\0&-2B\end{array}\right)_{p,q},
\end{eqnarray}
for 4=2$\times$2 possibilities corresponding to
\begin{eqnarray}
\left(\begin{array}{cc}h_{n-1}=h_n=B,&h_{n-1}=-h_n=B\\-h_{n-1}=h_n=B,&h_{n-1}=h_n=-B\end{array}\right)
\end{eqnarray}
and we denote the four-dimensional local Hilbert space as
\begin{eqnarray}
|\phi\rangle=|s=0,s^z=0\rangle;
|\psi_m\rangle=|s=1,s^z=m\rangle\,\,\,(m\in\{0,\pm1\}).
\end{eqnarray}
Since $H_0+H'(B)$ preserves U$(1)_z$,
only $|\phi\rangle$ and $|\psi_0\rangle$
can be mixed with each other.

We simply separate
several special cases:
\begin{itemize}
\item $0\leq B<1$

The four ground states corresponding to the above four possibilities can be put into a $2\times2$ matrix:
\begin{eqnarray}\label{asinglet}
(\Psi_n)_{p,q}=\frac{1}{\sqrt{1+\frac{a_{p,q}^2}{(1+\sqrt{a_{p,q}^2+1})^2}}}\left(|\phi\rangle+\frac{-a_{p,q}}{(1+\sqrt{a_{p,q}^2+1})^2}|\psi_0\rangle\right),\,\,\,(p,q\in\{1,2\}).
\end{eqnarray}

\item $B>1$
\begin{eqnarray}
(\Psi_n)_{2,1}=|\psi_+\rangle,\,\,(\Psi_n)_{1,2}=|\psi_-\rangle,
\end{eqnarray}
while $(\Psi_n)_{1,1}$ and $(\Psi_n)_{2,2}$ still has the same expression as in Eq.~(\ref{asinglet}).

\end{itemize}

Therefore,
\begin{eqnarray}
\langle\otimes_nO_n\rangle=\frac{1}{2^N}\text{Tr}\prod_{n=1}^N\langle\Psi_{n}|O_n|\Psi_{n}\rangle,
\end{eqnarray}
where the matrix
\begin{eqnarray}
\left(\langle\Psi_{n}|O_n|\Psi_{n}\rangle\right)_{p,q}\equiv\langle\left(\Psi_{n}\right)_{p,q}|O_n|\left(\Psi_{n}\right)_{p,q}\rangle
\end{eqnarray}

\subsection{Average value of the twisting operator}
Taking
\begin{eqnarray}
O_n=\exp\left\{\frac{2\pi i}{L}[2nS^z_{2n}+(2n+1)S^z_{2n+1}]\right\},
\end{eqnarray}
then,
due to the difference at the boundary,
\begin{eqnarray}\label{U_O_relation}
U=(-1)\prod_{n=1}^NO_n.
\end{eqnarray}
we obtain that
\begin{itemize}
\item $0\leq B<1$
\begin{eqnarray}
\langle\Psi_{n}|O_n|\Psi_{n}\rangle&=&\langle\Psi_{n}|\exp\left(\frac{2\pi i}{L}S^z_{2n+1}\right)|\Psi_{n}\rangle\nonumber\\
&=&\left(\begin{array}{cc}\cos\frac{\pi}{L}+i\Delta_B\sin\frac{\pi}{L}&\cos\frac{\pi}{L}\\\cos\frac{\pi}{L}&\cos\frac{\pi}{L}-i\Delta_B\sin\frac{\pi}{L}\end{array}\right),
\end{eqnarray}
since $S^z_{2n}+S^z_{2n+1}=0$,
where $\Delta_B\equiv\frac{4B}{1+\sqrt{1+4B^2}+\frac{4B^2}{1+\sqrt{1+4B^2}}}$,
and we have also used the fact that
\begin{eqnarray}
&&S^z_{2n+1}=-\frac{\tau_x}{2},\\
&&\exp\left(\frac{2\pi i}{L}\frac{-\tau_x}{2}\right)=\cos\frac{\pi}{L}-i\tau_x\sin\frac{\pi}{L},
\end{eqnarray}
under the two-dimensional basis $\{|\phi\rangle,|\psi_0\rangle\}$.

Diagonalizing the $n$-independent $\langle\Psi_{n}|O_n|\Psi_{n}\rangle$ gives the analytical result
\begin{eqnarray}
\langle U\rangle&=&(-1)\sum_{\eta=\pm1}\left[\frac{1}{2}\left(\cos\frac{\pi}{L}+\eta\sqrt{\cos^2\frac{\pi}{L}-\Delta_B^2\sin^2\frac{\pi}{L}}\right)\right]^{L/2}\nonumber\\
&=&-1+\frac{\pi^2}{8L}(2+\Delta_B^2)+\mathcal{O}(1/L^2).
\end{eqnarray}

\item $B>1$
\begin{eqnarray}
\langle\Psi_{n.}|O_n|\Psi_{n}\rangle&=&\langle\Psi_{n.}|\exp\left(\frac{2\pi i}{L}S^z_{2n+1}\right)|\Psi_{n}\rangle\nonumber\\
&=&\left(\begin{array}{cc}\cos\frac{\pi}{L}+i\Delta_B\sin\frac{\pi}{L}&\exp\left[\frac{-2\pi i}{L}\left(2n+\frac{1}{2}\right)\right]\\\exp\left[\frac{2\pi i}{L}\left(2n+\frac{1}{2}\right)\right]&\cos\frac{\pi}{L}-i\Delta_B\sin\frac{\pi}{L}\end{array}\right).
\end{eqnarray}
Thus,
after using the identity to eliminate the $n$-dependence within each multiplier
\begin{eqnarray}\label{identity}
\text{Tr}\prod_{n=1}^{N}\left(\begin{array}{cc}x+iy&z\exp(-in\theta)\\z\exp(in\theta)&x-iy\end{array}\right)=\text{Tr}\left\{\left[\exp(-i\frac{\sigma_z}{2}\theta)\left(\begin{array}{cc}x+iy&z\\z&x-iy\end{array}\right)\right]^N\exp\left(i\frac{\sigma_z}{2}N\theta\right)\right\},\nonumber\\
\end{eqnarray}
we obtain the analytical result
\begin{eqnarray}
&&\langle U\rangle\nonumber\\
&=&\sum_{\eta=\pm1}\left[\frac{1}{2}\left(\cos\frac{\pi}{L}\cos\frac{2\pi}{L}-\Delta_B\sin\frac{\pi}{L}\sin\frac{2\pi}{L}+\eta\sqrt{1-\left(\cos\frac{\pi}{L}\sin\frac{2\pi}{L}+\Delta_B\sin\frac{\pi}{L}\cos\frac{2\pi}{L}\right)}\right)\right]^{L/2}\nonumber\\
&=&1-\frac{\pi^2}{8L}(\Delta_B^2+8\Delta_B+9)+\mathcal{O}(1/L^2),
\end{eqnarray}
where the ``$(-1)$'' in Eq.~(\ref{U_O_relation}) is cancelled by $\exp(iN\theta\sigma_z/2)$ in Eq.~(\ref{identity}) because $\theta=4\pi/L=2\pi/N$.
\end{itemize}

\subsection{Calculation of staggering magnetism}
Due to the weak translation symmetry,
\begin{eqnarray}
\langle M_s\rangle=\langle-4S^z_2S^z_3\rangle.
\end{eqnarray}
We notice that under the $s^z=0$ basis $\{|\phi\rangle,|\psi_0\rangle\}$,
$-4S^z_2S^z_3=\left(\begin{array}{cc}1&1\\1&1\end{array}\right)$.
\begin{itemize}
\item $0\leq B<1$
\begin{eqnarray}
\langle M_s\rangle=\frac{1}{2^{N}}\text{Tr}\left(\begin{array}{cc}1&1\\1&1\end{array}\right)^{N}=1.
\end{eqnarray}

\item $B>1$
\begin{eqnarray}
\langle M_s\rangle=\frac{1}{2^{N}}\text{Tr}\left[\left(\begin{array}{cc}1&-1\\-1&1\end{array}\right)\left(\begin{array}{cc}1&1\\1&1\end{array}\right)^{N-1}\right]=0.
\end{eqnarray}

\end{itemize}

The above calculations show that the transition at $B_c=1$ is of first order.
Thus,
$B=B_c$ is a singular point leading to discontinuous jumps of various quantities.

These two phases correspond to two average SPT (ASPT) phases protected by the $U(1)_z \rtimes \mathbb{Z}_2^x$ symmetry \cite{ma2025topological}, which generalizes the notion of SPT phases to mixed states. 
The classification is given by the group cohomology $H^2(U(1)_z, H^1(\mathbb{Z}_2, \mathbb{Z}_2^x)) = \mathbb{Z}_2$. 
In the case of the onsite $U(1)_z \rtimes \mathbb{Z}_2^x$ symmetry,
the corresponding definition of ``cell'' will define the concept of ``trivial'' and ``nontrivial'' phases as follows.
Each cell is required to form a $U(1)_z \rtimes \mathbb{Z}_2^x$ representation,
so it must consist of even number of spin-1/2's.
There are two typical definitions of the cell:
(a) each cell by $(2n)$ and $(2n+1)$ sites;
(b) each cell by $(2n+1)$ and $(2n+2)$ sites.
In the example above, at $B=0$ the system is in a pure state as a trivial $U(1)_z \rtimes \mathbb{Z}_2^x$ ASPT phase under the definition~(a) since it is a product state while nontrivial under the definition~(b).
By contrast, at $B=\infty$ the system reduces to a product mixed state under the definition~(b) thereby belonging to the trivial ASPT phase.
However,
under the definition~(a) and its different topological order parameter $\langle U\rangle$ from $B=0$,
it must belong to the nontrivial ASPT phase.
Therefore,
for our current lattice models without any interest in defining the artificial choice of the cell,
only the distinction of topological phase is practically essential while which one is trivial or nontrivial is not substantial.

Here we remark that the mixed state at $B=\infty$ indeed satisfies a quantum information definition of ASPT in Ref. \cite{ma2025topological} where an ASPT state is defined as a symmetrically invertible gapped state. First, the “gapped’’ property of $\rho(B=\infty)=\otimes_{n}\frac{1}{2}\left(|\uparrow\downarrow\rangle\langle \uparrow\downarrow|+(|\downarrow\uparrow\rangle\langle \downarrow\uparrow|\right)_{2n-1,2n}$ follows from the fact that its conditional mutual information vanishes:
\[
I(A,B,C)=S_{AB}+S_{BC}-S_B-S_{ABC}=0
\]
for any tripartition $A,B,C$ of the chain and $S_M$ is the entropy of the reduced density matrix in the region $M$. Thus the state contains no long-range correlations in the quantum-information sense. Second, symmetric invertibility requires the existence of an auxiliary mixed state $\tilde{\rho}$ on an auxiliary Hilbert space such that one can define a diagonal onsite $U(1)_z\rtimes \mathbb{Z}^x_2$ generated by $\bm{S}^z_\text{tot}\otimes \tilde{\bm{S}^z}_\text{tot}$ and $R^{\pi}_{\pi}\otimes \tilde{R}^{\pi}_{\pi} $ where the tilde operators satisfy the same group relations. The condition is that $\rho\otimes \tilde{\rho}$ is two way connected to a pure product through
some locally symmetric finite-depth channels. For $\rho(B=\infty)$, we can show this property by considering two copies: $\rho(B=\infty)\otimes \rho(B=\infty)$ and apply the one-layer unitary circuit:
\[
U_1=\prod_{n}\frac{1+\vec{\sigma}_{2n}\cdot\vec{\tilde{\sigma}}_{2n-1}}{2}
\]
which swaps the physical spin at site $2n$ site and the auxiliary spin on $2n-1$ site. This unitary transformation preserves the diagonal $U(1)_z\rtimes \Z^x_2$ symmetry. After this transformation, the state becomes a product state state $\rho'=\otimes_{n}\rho'_n=\otimes_{n}\frac{1}{2}\left(|\uparrow\tilde{\downarrow}\rangle\langle \uparrow\tilde{\downarrow}|+(|\downarrow\tilde{\uparrow}\rangle\langle \downarrow\tilde{\uparrow}|\right)_{n}$ for either choice (a) or (b) of the unit cell. On each site $n$, this mixed state is two way connected to the single state $|\psi_{-}\rangle=\frac{1}{\sqrt{2}}(|\uparrow\tilde{\downarrow}\rangle-|\downarrow\tilde{\uparrow}\rangle)$ via symmetric finite-depth channels. Explicitly,
\[
\begin{split}
&\mathcal{N}_{12}[|\psi_-\rangle\langle\psi_-|]=\rho'_n, K^{12}_0=I, ~K^{12}_1=\sigma^z,\\
&\mathcal{N}_{21}[\rho'_n]=|\psi_-\rangle\langle\psi_-|, K^{21}_0=|\psi_-\rangle\langle \uparrow\tilde{\downarrow}|, ~K^{21}_1=|\psi_-\rangle\langle \downarrow\tilde{\uparrow}|, K^{21}_2=|\uparrow\tilde{\uparrow}\rangle\langle \uparrow\tilde{\uparrow}|, K^{21}_3=|\downarrow\tilde{\downarrow}\rangle\langle \downarrow\tilde{\downarrow}|
\end{split}
\]
where $K$ are the corresponding Kraus operators. Thus  $\rho'$ i locally-symmetrically two-way connected to a product of singlet state a product of singlet states, which establishes the symmetric invertibility of $\rho(B=\infty)$. We therefore include that $\rho(B=\infty)$ realizes an ASPT state in the sense of Ref. \cite{ma2025topological}.

In conclusion, we have the statement: If $\rho$ belongs to a $[U(1)_z\rtimes\mathbb{Z}^x_2]$ ASPT phase with strong $U(1)_z$ and weak $\mathbb{Z}_2^x$ symmetry, then $\text{Tr}(\rho U)=\pm1+O(1/L)$ and its thermodynamic limit identifies two distinct ASPT phases.

\section{ Spin–spin correlation in chiral  scalar triple-product model }
We consider chiral scalar triple-product spin model
\begin{equation}
\mathcal{H} _\text{ch} = \sum_{r}(-1)^i J\vec{S}_i\cdot(\vec{S}_{i+1}\times\vec{S}_{i+2}).
\label{eq:CTP}
\end{equation}
with ground state $|\text{G.S.}\rangle$ and the channel
\begin{eqnarray}
    \mathcal{N}=\otimes_i \mathcal{N}^z_i, \quad \mathcal{N}^z_i[\rho]=(1-p)\rho+4 p S^z_i\rho S^z_i
\end{eqnarray}

When $p=0$,
i.e., $\mathcal{N}[|\text{G.S.}\rangle\langle\text{G.S.}|]=|\text{G.S.}\rangle\langle\text{G.S.}|$,
an iDMRG study \cite{PhysRevB.99.205121} shows that it exhibits critical behavior which belongs to the same universality class as the spin-1/2 Heisenberg chain. In particular, the spin–spin correlation function decays as a power law
\begin{equation}
    \langle\text{G.S.}| S^z_i S^z_j|\text{G.S.}\rangle \sim \frac{(\log |i-j|)^{0.5}}{|i-j|}
\end{equation}
When the channel is included with nonzero $p$, the spin–spin correlation function remains invariant:
\begin{equation}\label{eq:correlation}
   \text{Tr}( \mathcal{N}[|\text{G.S.}\rangle\langle\text{G.S.}|]S^z_i S^z_j)= \text{Tr}( \mathcal{N}[|\text{G.S.}\rangle\langle\text{G.S.}|S^z_i S^z_j])=\langle\text{G.S.}| S^z_i S^z_j|\text{G.S.}\rangle
\end{equation}
where we use the fact that the channel commutes with $S^z_i$. Hence, the spin–spin correlation function always exhibits power-law decay, independent of $p$.

For such a mixed state, we can prove that it cannot be weak $\mathbb{Z}_2^x$-mSRE by contradiction without invoking {\bf Corollary~7} as follows.
Suppose that
\begin{equation}
     \mathcal{N}[|\text{G.S.}\rangle\langle\text{G.S.}|]=\text{Tr}_E\{D_{S\cup E}\left[\rho_0\otimes|0_E\rangle\langle 0_E|\right]D_{S\cup E}^\dagger\}
\end{equation}
Then, if the distance between $i$ and $j$ is sufficiently large, we obtain
\begin{equation}
\begin{split}
     &\text{Tr}_{S}( \mathcal{N}[|\text{G.S.}\rangle\langle\text{G.S.}|]S^z_i S^z_j)
     \\
     =&\text{Tr}_{SE}\{D_{S\cup E}\left[\rho_0\otimes|0_E\rangle\langle 0_E|\right]D_{S\cup E}^\dagger S^z_i S^z_j\}
     \\=&\text{Tr}_{SE}\{\left[\rho_0\otimes|0_E\rangle\langle 0_E|\right] \Tilde{S}^z_i \Tilde{S}^z_j\}
     \\=&\text{Tr}_{SE}\{\left[\rho_0\otimes|0_E\rangle\langle 0_E|\right] \Tilde{S}^z_i\}\text{Tr}_{SE}\{\left[\rho_0\otimes|0_E\rangle\langle 0_E|\right] \Tilde{S}^z_j\}
     \\=&0
\end{split}
\end{equation}
where $\Tilde{S}^z_i$ is $D_{S\cup E}^\dagger S^z_i D_{S\cup E}$ is a local operator that is odd under $R^x_{\pi} \otimes R_E$. Here we also use the fact that $\rho_0$ is a product of weak $\mathbb{Z}^x_2$-symmetric mixed states. This result contradicts with the power-law decaying behavior in Eq.\eqref{eq:correlation}, which completes the proof.
\end{document}